\documentclass[twocolumn,nolinenumbers]{aastex631}
\usepackage{graphicx}
\usepackage{amsmath}	% Advanced maths commands
\usepackage{amssymb}	% Extra maths symbols
\usepackage{gensymb}
\usepackage{cancel}
\usepackage{enumerate}

\newcommand{\beq}{\begin{equation}}
\newcommand{\eeq}{\end{equation}}

\newcommand{\vA}{\boldsymbol{v}_{\rm A}}

\newcommand{\vB}{\boldsymbol{B}}

\newcommand{\vu}{\boldsymbol{u}}
\newcommand{\vb}{\boldsymbol{b}}

\newcommand{\pd}{\partial}
\newcommand{\pddt}[1]{\frac{\pd#1}{\pd t}}
\newcommand{\pddtau}[1]{\frac{\pd#1}{\pd \tau}}
\newcommand{\ada}{\frac{\dot{a}}{a}}

\newcommand{\avg}[1]{\overline{#1}}
\newcommand{\fluc}[1]{\widetilde{#1}}
\newcommand{\vp}{\boldsymbol{p}}

\newcommand{\vAx}{v_{\mathrm{A}x}}
\newcommand{\vAy}{v_{\mathrm{A}y}}

\newcommand{\vpu}{\vp \cdot \vu_1}
\newcommand{\vpul}{\vp\cdot\vu_{1\lambda}}
\newcommand{\vap}{\vp\cdot\vA}
\newcommand{\vAt}{\boldsymbol{v}_{\rm A T}}

\newcommand{\vn}{\boldsymbol{n}}

\newcommand{\ucomp}{u_{1\mathrm{comp}}}
\newcommand{\ucompl}{u_{1\mathrm{comp}\lambda}}
\newcommand{\dbsqm}{\delta (b^2)_{1\lambda}}
\newcommand{\dbsq}{\delta (b^2)_1}
\newcommand{\vAm}{v_{\mathrm{A}m}}
\newcommand{\btr}{b_{\rm T}}

\shorttitle{Large-amplitude Alfv\'en waves in the expanding solar wind}
\shortauthors{Mallet et al.}

\begin{document}

\title{Evolution of large-amplitude Alfv\'en waves and generation of switchbacks in the expanding solar wind}

\author{Alfred Mallet}
\affil{Space Sciences Laboratory, University of California, Berkeley CA 94720, USA}

\author{Jonathan Squire}
\affil{Physics Department, University of Otago, Dunedin 9010, New Zealand}

\author{Benjamin D. G. Chandran}
\affil{Space Science Center, University of New Hampshire, Durham, NH 03824, USA}

\author{Trevor Bowen}
\affil{Space Sciences Laboratory, University of California, Berkeley CA 94720, USA}

\author{Stuart D. Bale}
\affil{Space Sciences Laboratory, University of California, Berkeley CA 94720, USA}
\affil{Physics Department, University of California, Berkeley CA 94720, USA}

\correspondingauthor{Alfred Mallet}
\email{alfred.mallet@berkeley.edu}

\date{\today}

\begin{abstract}
Motivated by recent Parker Solar Probe (PSP) observations of ``switchbacks" (abrupt, large-amplitude reversals in the radial magnetic field, which exhibit Alfv\'enic correlations) we examine the dynamics of large-amplitude Alfv\'en waves in the expanding solar wind. We develop an analytic model which makes several predictions: switchbacks should preferentially occur in regions where the solar wind plasma has undergone a greater expansion, the switchback fraction at radii comparable to PSP should be an increasing function of radius, and switchbacks should have their gradients preferentially perpendicular to the mean magnetic field direction. The expansion of the plasma generates small compressive components as part of the wave's nonlinear evolution: these are maximized when the normalized fluctuation amplitude is comparable to $\sin\theta$, where $\theta$ is the angle between the propagation direction and the mean magnetic field. These compressive components steepen the primary Alfv\'enic waveform, keeping the solution in a state of nearly constant magnetic field strength as its normalized amplitude $\delta B/B$ grows due to expansion. The small fluctuations in the magnetic-field-strength are minimized at a particular $\theta$-dependent value of $\beta$, usually of order unity, and the density and magnetic-field-strength fluctuations can be correlated or anticorrelated depending on $\beta$ and $\theta$. Example solutions of our dynamical equation are presented; some do indeed form magnetic-field reversals. Our predictions appear to match some previously unexplained phenomena in observations and numerical simulations, providing evidence that the observed switchbacks result from the nonlinear evolution of the initially small-amplitude Alfv\'en waves already known to be present at the coronal base.
\end{abstract}

\section{Introduction}
Alfv\'en waves are ubiquitous in the solar wind and corona \citep{belcher1971}, and are considered to be the basic building blocks of the turbulence that heats the corona and accelerates the solar wind. As the solar wind plasma travels away from the sun and expands, Alfv\'en waves carried along with the plasma decay in amplitude \citep{parker1965}. Flux conservation means that the overall (mean) radial magnetic field strength $\avg{B}_R$ has a somewhat faster decay than the fluctuations: the combination of these two laws mean that the \emph{normalized} amplitude of Alfv\'en waves, $\delta B/\avg{B}_R$, increases as the solar wind plasma travels away from the sun. Even when the nonlinear effects of turbulence are taken into account \citep{cranmer2005,verdini2007,ballegooijen2016,ballegooijen2017,perezchandran2013,chandran2019}, initially low-amplitude waves eventually attain a large normalized amplitude. In this paper, we study the dynamical evolution of Alfv\'en waves of arbitrary amplitude in the expanding solar wind.

Additional motivation for the study of large-amplitude Alfv\'en waves comes from the striking observation of ``switchbacks": large-amplitude, abrupt reversals of the magnetic field, observed by NASA's Parker Solar Probe (PSP) mission in the inner solar wind and corona \citep{bale2019,horbury2020,dudokdewit2020,krasnoselskikh2020,farrell2020,mcmanus2020,mozer2020,laker2020}. These observations have prompted several attempts at theoretical explanation, with different theories suggesting that they may be the result of coronal reconnection forming either flux ropes \citep{drake2020} or complex fast-mode-like structures \citep{zank2020}, the nonlinear Kelvin-Helmholtz instability \citep{ruffolo2020}, or differential flow between different solar wind sources \citep{schwadron2021}. An alternative, potentially much simpler hypothesis is that switchbacks may form naturally \emph{in situ} from initially low-amplitude Alfv\'en waves (AWs) \citep{squire2020} due to the growth of their amplitude relative to the mean magnetic field in the expanding Solar Wind \citep{grappin1996}. These Alfv\'en waves are known to be ubiquitous in the corona already \citep{depontieu2007,ballegooijen2016}, and so if it works, this model does not require any ``special event" to create the switchbacks.

Let us separate the magnetic field and velocity into their mean and fluctuations, $\vB = \avg{\vB}+\delta \vB$, $\vu = \avg{\vu}+\delta\vu$. In a homogeneous medium (constant mean velocity $\avg{\vu}$ and magnetic field $\avg{\vB}$), any configuration of magnetic field and velocity fluctuations with an Alfv\'enic correlation between its velocity and magnetic field fluctuations,
\beq
\delta \vu = \pm \delta \vB/\sqrt{4\pi\rho},\label{eq:du}
\eeq
which additionally has constant magnetic-field strength, density, and pressure,
\begin{align}
B^2 = \text{const.},\label{eq:B2const}\\
\rho=\text{const.},\label{eq:rhoconst}\\
P=\text{const.},\label{eq:Pconst}
\end{align}
is an exact nonlinear solution to the MHD equations \citep{goldstein1974}, propagating with group velocity equal to
\beq
\boldsymbol{v}_g=\pm\vA=\pm\avg{\vB}/\sqrt{4\pi\rho},\label{eq:vA}
\eeq
the Alfv\'en velocity (in the frame in which $\avg{\vu}=0$), regardless of its amplitude; i.e. without steepening. This property is unique amongst the large-amplitude MHD waves. The properties (\ref{eq:du}--\ref{eq:vA}) comprise the definition of a large-amplitude Alfv\'en wave in a homogeneous medium; we will also refer to this as a \emph{perfect Alfv\'en wave}, and/or say that a configuration is \emph{Alfv\'enic} if it satisfies (or nearly satisfies) (\ref{eq:du}--\ref{eq:Pconst}). More often in the literature, only (\ref{eq:du}) is used to define the Alfv\'enic state, but the others are also necessary for a configuration to propagate at $\vA$, and are crucial for large-amplitude waves. Due to (\ref{eq:B2const}), the magnetic-field vector moves on the surface of a sphere; thus, this configuration is also referred to as a ``spherically-polarized Alfv\'en wave". Moreover, (\ref{eq:B2const}) implies that, unlike the linearized Alfv\'en wave, the large-amplitude Alfv\'en wave has a parallel (to $\avg{\vB}$) component $\delta B_\parallel$. This constraint also means that a monochromatic wave is impossible, since (\ref{eq:B2const}) involves both terms quadratic and linear in the fluctuations (\citealt{barnes1974}, see also Sec.~\ref{sec:b0xsc}).

\begin{deluxetable*}{|l|c|l|}
\tablehead{
\colhead{Prediction} & \colhead{Section} & \colhead{Observational evidence / \emph{possible test}} }
\tablecaption{Predictions of the model, their location in the paper, and either existing observational evidence or a suggested observational test to confirm or falsify the model.\label{tab:pred}}
\startdata
Nearly-constant magnetic field strength \citep[cf.][]{barnes1974}.& \ref{sec:sbeq} & Agrees with \citet{horbury2020} (among many others).\\
\hline
Larger amplitude with more expansion. \citep[cf.][]{hollweg1974}& \ref{sec:ampsc} & \emph{Switchbacks should occur in ``patches" where the plasma} \\
& & \emph{has undergone more expansion \citep{bale2021inprep}.}\\
\hline
 Increasing switchback fraction with radial distance from the Sun. &\ref{sec:ampsc} & Agrees with \citet{mozer2020,badman2020b}.\\
 \hline
Angle-dependent threshold for radial field reversals: preference for & \ref{sec:b0xsc} & Agrees with \citet{laker2020}: switchbacks elongated\\
perpendicular structures. & &along $\avg{B}$. Agrees with initial-condition dependence of \\
& & switchback fraction in simulations \citep{squire2020}.\\
\hline
 Compressive components: scaling with expansion rate. & \ref{sec:vpusc} & \emph{Measure compressions as a function of expansion rate} \\
& & \emph{(cf. slow/fast wind as high/low expansion surrogates).}\\
\hline
Magnetic compressibility minimized for $\beta\sim1$. & \ref{sec:b2dsc} & Agrees with simulations of turbulence \citep{squire2020}. \\
\hline
Angle-dependent transition in polarization state $\xi=(\delta|\vB|/\avg{B})/(\delta\rho/\avg{\rho})$. & \ref{sec:polsc} & \emph{Could be compared with \citet{larosa2020}.} \\
\hline
\enddata
\end{deluxetable*}
\vskip-0.9cm
Rather strikingly, all of (\ref{eq:du}--\ref{eq:vA}) are (usually) nearly satisfied by many switchbacks \citep{bale2019,kasper2019,horbury2020}\footnote{Recent evidence suggests, however, that there is a small but systematic variation in the compressive components (\ref{eq:B2const}--\ref{eq:Pconst}) \citep{farrell2020,larosa2020}, a point to which we will return later.}. This provides some simple but powerful evidence in support of the \emph{in situ} Alfv\'en wave generation hypothesis: the \citet{squire2020} model is currently the only one that naturally includes these basic correlations. Other models must invoke some (largely unspecified) process of ``Alfv\'enization" \citep{drake2020,ruffolo2020,schwadron2021} to explain the constant-magnetic-field-strength observation. Moreover, \citet{squire2020} performed numerical simulations of expanding Alfv\'enic turbulence, and showed that the initial condition of random, small-amplitude, outward-propagating Alfv\'en waves, upon expansion, does indeed naturally produce switchbacks. \citet{shoda2021} also found this using compressible MHD simulations in a narrow magnetic flux tube extending from the coronal base out to $40 R_\odot$. 

In an inhomogeneous medium, it can be shown that the overall root-mean-square (RMS) amplitude of a large-amplitude Alfv\'en wave obeys the same law as in the small-amplitude case \citep[][see also Sec.~\ref{sec:ampsc}]{hollweg1974,barnes1974}. However, because $B^2= (\delta\vB + \avg{\vB})^2$, and $\delta B/\avg{B}$ grows with expansion, small variations in $B^2$ are constantly being produced, and thus the expansion constantly pushes the wave out of its perfect constant-$B^2$ state. How is it then possible to maintain the Alfv\'enic nature of the wave over long timescales? In other words, while other models of switchback formation have problems explaining why the switchbacks \emph{become} Alfv\'enic, an important question to ask of the \citet{squire2020} Alfv\'en wave expansion model is how the switchbacks \emph{remain} so Alfv\'enic, given the apparently harmful effects of expansion; this is the focus of the current paper. We attack the problem in a simplified fashion: we neglect every effect except a simple spherical expansion, and adopt  the (locally-isothermal) Expanding Box MHD model \citep[][henceforth EBM]{velli1992,grappin1993}. We apply a two-timescale analysis (the details of which are in the next section) to arbitrary-amplitude Alfv\'enic fluctuations that, at any instant in time, vary along a single spatial direction. We show that the waveform of the primary Alfv\'enic component obeys a relatively simple dynamical equation (Eq.~\ref{eq:sbeqmain}) on the slow expansion timescale. Physically, the expansion drives a small non-constant component of the magnetic pressure $B^2$, which then drives a compressive flow, distorting the waveform.

The utility of our dynamical equation, introduced in Section \ref{sec:sbeq}, is that it allows us to make a number of detailed predictions that go beyond the basic properties of large-amplitude Alfv\'en waves (\ref{eq:du}--\ref{eq:vA}); these are described in Section \ref{sec:scal}. We re-derive \citeauthor{hollweg1974}'s \citeyear{hollweg1974} result on the scaling of the amplitude of large-amplitude Alfv\'en waves with expansion, and thus predict that switchbacks should occur preferentially in patches of solar wind where the super-radial expansion of field lines and plasma is large (Sec.~\ref{sec:ampsc}). We also argue that this amplitude evolution means that at radial distances $R$ comparable to the perihelia of PSP, the fraction of the volume filled with fluctuations that reverse the radial field should be an increasing function of $R$, agreeing with the observations of \citet{mozer2020} and \citet{badman2020b}.
Using a simple geometrical argument, we also show that it is easier for fluctuations to reverse the mean magnetic field and form switchbacks when the direction of variation (wavevectors) is nearly perpendicular to $\avg{\vB}$ (Sec.~\ref{sec:b0xsc}) -- this may partly explain the observations of \citet{laker2020} that switchbacks are highly elongated along the mean magnetic field direction. A similar anisotropy of the switchbacks was observed in the numerical turbulence simulations of \citet{squire2020}, who also observed that the switchback fraction was significantly higher in runs for which there were more extremely perpendicular modes in the initial condition\footnote{In particular, their Gaussian initial conditions had more power in modes with large $k_\perp$ initially.}. 
We outline the scaling properties of the new expansion-driven compressible components of the wave, and discuss how this is related to nonlinear steepening and distortion (Sec.~\ref{sec:vpusc}).  
In Sec.~\ref{sec:b2dsc}, we show that the small variation in magnetic-field strength introduced by expansion is minimized at $\beta\sim1$, which agrees with the turbulence simulations of \citet{squire2020}. 
Sec.~\ref{sec:polsc} shows further that whether the small variations in $B^2$ and $\rho$ are correlated or anti-correlated (``fast-wave-like" or ``slow-wave-like" respectively) depends on the propagation angle and $\beta$, potentially explaining the fact that both these behaviours are seen in the observational data \citep{larosa2020}. 
For clarity, all of these predictions are summarized in Table \ref{tab:pred}. 

Having studied the general properties of our equation in some detail, we then present in Section \ref{sec:evol} some sample numerical solutions of our dynamical equation with different initial conditions, examining the conditions which generically give rise to switchbacks and confirming the results of our scaling analyses.

\section{Derivation of the switchback equation}\label{sec:sbeq}
In this Section, we will derive an equation governing the evolution of the waveform of an Alfv\'enic structure of arbitrary amplitude as it travels away from the Sun and expands. If the reader prefers to skip the details, a summary of our results is given in Sec.~\ref{sec:derivsumm}. Before we begin, we will briefly outline the major assumptions and restrictions of our analysis. 

First, our calculation uses the Expanding Box model (EBM, \citealt{velli1992,grappin1993}), which expresses the MHD equations in a small parcel of plasma, in a frame comoving with the mean solar wind velocity $U$ (a constant in the EBM), undergoing spherical expansion (in the $y$- and $z$-directions) as it moves radially (in the $x$-direction) away from the sun. While constant $U$ and spherical expansion are undoubtedly rather poor approximations to the corona, this model contains the important physical mechanism we are interested in: expansion makes $\avg{\vB}$ and $\delta \vB$ decay differently, thus increasing the normalized amplitude and also pushing the wave out of perfectly constant $B^2$. Thus, we feel that the EBM is an acceptable minimal model with which to study the dynamics of large-amplitude Alfv\'en waves as they travel away from the sun. Moreover, the way it replaces all of the inhomogeneity in space with a time-dependent background (in the Expanding Box frame) will prove analytically very useful. In our analysis, we assume that the timescale of the Alfv\'enic fluctuations, $T_{\rm A}$, is much shorter than the timescale associated with the expansion, $T_{\rm exp}$. We then define the small parameter
\beq
\epsilon = \frac{T_{\rm A}}{T_{\rm exp}},\label{eq:epsrestrict}
\eeq
and expand the solution to the EBM equations in powers of $\epsilon$. We restrict ourselves to solutions for which the leading-order solutions for $B^2$, $\rho$, and $P$ are constant (cf. Eqs.~\ref{eq:B2const}--\ref{eq:Pconst}). We additionally assume that in the initial condition, the solution varies only along a single spatial direction: the expansion of the box subsequently causes that direction to rotate with time, but at each instant, the solution is one-dimensional. We will repeat these major assumptions at the points at which each is introduced in the derivation.

As mentioned in the introduction, our assumption that $B^2$, $\rho$, and $P$ are constant to leading order is rather well motivated by the observational evidence, and means that our solutions behave at zeroth order like large-amplitude Alfv\'en waves. In addition, the smallness of $\epsilon$ does appear to be somewhat satisfied for the real switchbacks: \citet{laker2020} found that the parallel length of the switchback is around $1R_\odot$, meaning their Alfv\'en time is $T_A\sim R_\odot/v_A$, while the expansion time around the radial location at which they are observed is $T_{\rm exp}\sim 30R_\odot/V_{\rm SW}$; using $v_A\approx 100\mathrm{kms^{-1}}$ and $V_{\rm SW}\approx 300\mathrm{kms^{-1}}$, the ratio of these times is indeed a small parameter, $\epsilon \approx 1/10$. Our assumption of one-dimensional spatial variation, however, is more seriously objectionable: real switchbacks are clearly three-dimensional. This assumption limits the validity of our solution to regions where the direction of the gradients is very close to constant. 

Finally, let us point out two aspects of the solution that we do not restrict: we allow the zeroth order fluctuations of magnetic field and velocity to have arbitrary amplitude, and we permit any zeroth-order waveform for which $B^2$ is spatially constant.

\subsection{Preliminaries}
The starting point for our analysis is, as previously mentioned, the Expanding Box Model (EBM) \citep{velli1992,grappin1993}, which consists of the following equations:
\begin{align}
\pddt{\rho} + \nabla\cdot \left(\rho \vu\right) &= -2\ada\rho,\label{eq:cont}\\
\pddt{\vu}+\vu\cdot \nabla \vu &= -\frac{1}{\rho}\nabla\left(c_s^2\rho + \frac{B^2}{8\pi}\right)+\frac{\vB \cdot \nabla\vB}{4\pi\rho} - \ada \mathbb{T}\cdot \vu,\label{eq:mom}\\
\pddt{\vB} + \vu \cdot \nabla\vB &= \vB\cdot \nabla\vu - \vB\nabla\cdot\vu -\ada \mathbb{L}\cdot \vB,\label{eq:ind}\\
\nabla \cdot \vB &= 0,\label{eq:divb}
\end{align}
where $\mathbb{T}=\mathrm{diag}(0,1,1)$, $\mathbb{L}=\mathrm{diag}(2,1,1)$, and $\nabla = (\pd_x,a^{-1}\pd_y,a^{-1}\pd_z)$. The co-moving coordinates $\boldsymbol{r}=(x,y,z)$ do not change as the plasma travels away from the sun and expands; the corresponding coordinates in the inertial, non-expanding frame moving with the solar wind velocity (a.k.a. the ``solar wind frame") are $\boldsymbol{r}_g=(x_g,y_g,z_g) = (x,ay,az)$. The expansion factor $a = 1+\dot{a} t$, and $\dot{a}$ is a constant (because the solar wind velocity $U$ is assumed constant); the expansion timescale $T_{\rm exp}=a/\dot{a}$ (cf. Eq.~\ref{eq:epsrestrict}). We assume that the plasma is locally isothermal, $c_s^2 = c_s^2(t)$. In the present analysis we will not need to specify the precise time dependence; part of our solution will be a function of the instantaneous value of $\beta=c_s^2/v_A^2$.
We restrict ourselves to one-dimensional solutions that depend on the EBM spatial coordinates $\boldsymbol{r}=(x,y,z)$ only through the function 
\beq
\lambda = \vp^*\cdot\boldsymbol{r},
\eeq
where $\vp^*$ is constant in space and time; since the EBM coordinates are also constant in time, $\lambda$ is not a function of $t$. In the inertial solar wind frame, these solutions vary along the time-dependent vector $\vp$, written in terms of $\vp^*$ as
\beq
\vp(t) = \nabla \lambda = \begin{pmatrix}
p^*_x\\
p^*_y/a\\
p^*_z/a
\end{pmatrix}.\label{eq:prot}
\eeq
We additionally assume (Eq.~\ref{eq:epsrestrict}) that fluctuations vary on both a short timescale $T_A$ and the long timescale $T_{\rm exp} = a/\dot{a}$, and take $\epsilon=\dot{a}T_A/a$ to be a small parameter. Specifically, we assume that a fluctuation $f$ can be written in the form
\beq
f = f(\lambda,t,\tau),
\eeq
where
\beq
\tau = \epsilon t.
\eeq
We will use the notation
\beq
f_\lambda = \left.\frac{\pd f}{\pd \lambda}\right|_{t,\tau}, \quad f_t=\left.\frac{\pd f}{\pd t}\right|_{\lambda,\tau}, \quad f_\tau=\left.\frac{\pd f}{\pd \tau}\right|_{\lambda,t},
\eeq
to denote partial derivatives with respect to each of the arguments. In this one-dimensional case, the gradient of any function $f$ may be written
\beq
\nabla f = \vp f_\lambda,
\eeq
We may write the partial time derivative at constant $\boldsymbol{r}$ as
\begin{align}
\left.\frac{\pd f}{\pd t}\right|_{\boldsymbol{r}} &= f_t + \frac{d \tau}{d t} f_\tau, \nonumber\\
&= f_t + \epsilon f_\tau.
\end{align}
We will additionally define
\beq
\alpha = \frac{\dot{a}}{\epsilon a}.\label{eq:alphadef}
\eeq

We proceed by separating all variables into their spatial mean and fluctuating components, e.g.
\beq
f = \avg{f} + \fluc{f},
\eeq
where
\beq
\avg{f}(t,\tau)=\frac{1}{\Lambda}\int_0^\Lambda f(\lambda,t,\tau) d\lambda,\label{eq:avgdef}
\eeq
and we require that the solution is periodic in $\lambda$ with period $\Lambda$.\footnote{It is also possible to choose as the boundary condition that all quantities $f\to \avg{f}$ as $\lambda\to\infty$; since it is more numerically convenient, we use periodic boundary conditions in the present analysis. When comparing our results to the real world, it is important to remember that the EBM is only valid if $\Lambda\ll pR$.} Note that the spatial mean of the fluctuating component vanishes by definition, $\avg{\fluc{f}}=0$.

We expand the fluctuations as follows:
\begin{align}
\fluc{\vB} = \vB_0 + \epsilon\vB_1 + \ldots,\\
\fluc{\vu} = \vu_0 + \epsilon\vu_1 + \ldots,\\
\fluc{\rho} = \epsilon\rho_1 + \ldots,\label{eq:ordering}
\end{align}
where 
\begin{align}
&B_0\sim B_1 \sim \ldots \sim B_n \sim \avg{B},\nonumber\\
&u_0\sim u_1 \sim \ldots \sim u_n \sim v_{\rm A},\nonumber\\
&\rho_1\sim\rho_2\sim\ldots \sim \rho_n \sim \avg{\rho},\label{eq:comparable}
\end{align}
i.e., the amplitude of the zeroth-order fluctuation in $\vB$, $B_0$, is comparable to $\avg{B}$. Because we would like to study Alfv\'en waves, $\rho_0 = 0$ and so $\fluc{\rho} \ll \avg{\rho}$. Similarly, we require the fluctuations in the magnetic field \emph{strength} to be small:
\beq
\frac{\fluc{B^2}}{B_0^2} \sim O(\epsilon).\label{eq:nearlyalf}
\eeq
Since, to first order,
\beq
B^2 = |\avg{\vB} + \vB_0 + \vB_1|^2 = |\avg{\vB} + \vB_0|^2 + 2\epsilon(\avg{\vB}+\vB_0)\cdot \vB_1,
\eeq
we must therefore have 
\beq
(\avg{\vB}+\vB_0)\cdot \vB_{0\lambda} = 0.\label{eq:funny}
\eeq
Eq.~(\ref{eq:divb}) may be written
\beq
\vp \cdot \vB_\lambda = 0,\label{eq:divbfirst}
\eeq
which upon integration over $\lambda$ at constant $t,\tau$ means that
\beq
\vp \cdot \vB = g(t,\tau).
\eeq
Taking the spatial mean of this equation, we find that $g(t,\tau) = \vp\cdot \avg{\vB}$, i.e.,
\beq
\vp \cdot \vB = \vp \cdot \avg{\vB}, \quad \vp \cdot \fluc{\vB} = 0.\label{eq:pb}
\eeq

\subsection{Zeroth order}
At this order, the expansion terms in (\ref{eq:cont}--\ref{eq:ind}) are absent since $\dot{a}/a\sim O(\epsilon)$, and the system reduces to the standard (one-dimensional) MHD equations. Writing this system in conservative form and taking the spatial mean, we find that, at zeroth order,
\beq
\avg{\rho}_t = 0, \quad
\avg{\vu}_t = 0, \quad
\avg{\vB}_t =0,\label{eq:meanzeroth}
\eeq
and so at this order, our analysis recovers the properties of large-amplitude Alfv\'en waves in a constant, homogeneous background \citep{barnes1974}. We take $\avg{\vu}=0$; this is clearly consistent with these zeroth-order equations, and we will check later that this is also the case at first order. 

Now, we subtract (\ref{eq:meanzeroth}) from the relevant zeroth-order equations to find the fluctuating parts. The fluctuating part of the continuity equation (\ref{eq:cont}) at zeroth order reads
\beq
\avg{\rho} \vp \cdot \vu_{0\lambda} = 0,
\eeq
and so we may take $\vp \cdot \vu_0 = 0$ (using the same reasoning as led from Eq.~\ref{eq:divbfirst} to Eq.~\ref{eq:pb}). The fluctuating parts of the zeroth-order momentum (\ref{eq:mom}) and induction (\ref{eq:ind}) equations yield
\begin{align}
\vu_{0t} &= \frac{\vp\cdot\avg{\vB}}{4\pi\avg{\rho}} \vB_{0\lambda},\label{eq:fluczeroth}\\
\vB_{0t} &= \vp\cdot\avg{\vB}\vu_{0\lambda},
\end{align}
whose solutions are
\beq
\vu_0=\vu_0(\lambda \mp \vap t, \tau) = \mp \frac{v_{\rm A}}{\avg{B}} \vB_0(\lambda \mp \vap t, \tau),\label{eq:vu0}
\eeq
where the constant of integration is zero because these are fluctuations. This solution propagates at group velocity $\pm \vA$ in the inertial solar wind frame, and the relationship between the velocity and magnetic-field fluctuations is Alfv\'enic; it is an Alfv\'en wave, as expected. We will choose the lower sign in Eq.~(\ref{eq:vu0}), and introduce Alfv\'en units for the magnetic field fluctuations,
\beq
\vb_0 = \frac{\vB_0}{\sqrt{4\pi\avg{\rho}}} = \vu_0.
\eeq
Finally, the zeroth-order solution shows that the Alfv\'en-wave timescale is $T_A \sim (\vap)^{-1}$; for concreteness, we define
\beq
T_A=(\vap)^{-1}.\label{eq:tadef}
\eeq
\subsection{First order}
To go further, we will look for a solution in which all fluctuating quantities depend on $\lambda$ and $t$ only via $\lambda + \vap t$ (as well as on $\tau$); i.e., a (slowly-evolving) wave. This also means that we may express the $t$-derivative in terms of the $\lambda$-derivative,
\beq
f_t = \vap f_\lambda,
\eeq
thus removing the fast time dependence from the equations.

At first order, the mean of the continuity equation gives
\beq
\avg{\rho}_\tau +2\alpha\avg{\rho}= -\avg{(\rho_1\vu_0)}_\lambda,
\eeq
where we have used the fact that the spatial mean of a fluctuation is zero. The term on the RHS vanishes since it is the mean of a total derivative; thus,
\beq
\avg{\rho} \propto a^{-2}.
\eeq
The fluctuating part of the continuity equation now gives
\beq
\vap \rho_{1\lambda} + \avg\rho \vpul=0,
\eeq
and so
\beq
\boxed{\frac{\rho_1}{\avg\rho} = - \frac{\vpu}{\vap},\label{eq:dens}}
\eeq
using the fact that fluctuations have zero spatial average. The mean momentum equation at first order gives
\beq
\avg{\vu}_\tau + \alpha \mathbb{T}\cdot\avg{\vu} =  -\vp\avg{(\vb_1\cdot\vb_0)}_\lambda-\avg{\vpu \vu_{0\lambda}} - \avg{\frac{\rho_1}{\rho}\vap \vb_{0\lambda}}.
\eeq
The first term on the RHS results from the average of the total-pressure-gradient term; since it is the mean of a total derivative, it is zero. Upon inserting (\ref{eq:dens}) and the zeroth-order solution (\ref{eq:vu0}), the second and third terms on the RHS cancel; thus $\avg{u}_x \propto a^0$, $\avg{u}_{y,z}\propto a^{-1}$, and if $\avg{\vu}=0$ initially, it will remain zero throughout the evolution. This shows that our assertion that $\avg{\vu}=0$ is also consistent at first order.

The fluctuating momentum equation at first order is
\begin{align}
{\vu}_{0\tau} + \vp \left[ c_s^2\frac{\rho_1}{\avg\rho}+ {\vb_1\cdot(\vA+\vb_0)}\right]_\lambda&+\alpha\mathbb{T}\cdot \vu_0 \nonumber \\& = \vap (\vb_1 - \vu_1)_\lambda,\label{eq:mom1}
\end{align}
where we have used (\ref{eq:dens}) to cancel some terms. 

To analyse the induction equation, notice the following relation between the $\tau$-derivative of $\vB$ and $\vb=\vB/\sqrt{4\pi\avg{\rho}}$:
\beq
{\vb}_\tau = \frac{\vB_{\tau}}{\sqrt{4\pi\avg\rho}}-\frac{\vb{\avg{\rho}_\tau}}{2\avg\rho} = \frac{{\vB_{\tau}}}{\sqrt{4\pi\avg\rho}}+\alpha\vb.
\eeq
Using this, the mean of the induction equation at first order is
\beq
\boldsymbol{v}_{{\rm A}\tau}+\alpha \hat{\boldsymbol{x}}\vAx = - \avg{\vpu\vb_{0\lambda}}-\avg{\vb_0\vpul}.
\eeq
The terms on the RHS together comprise the mean of a total derivative, and are therefore equal to zero; thus,
\beq
\vAx \propto a^{-1}, v_{{\rm A}y,z}\propto a^0.\label{eq:meanva}
\eeq
Using (\ref{eq:prot}) and (\ref{eq:meanva}), $\vap = T_A^{-1} \propto a^{-1}$. It then follows from (\ref{eq:epsrestrict}), (\ref{eq:alphadef}) and (\ref{eq:tadef}) that
\beq
\epsilon = \frac{T_A}{T_{\rm exp}} = \frac{\dot a}{a\vap} = \text{const.}, \quad \frac{\alpha}{\vap} = 1.\label{eq:epsconst}
\eeq
This relation will be useful in analysing the scaling properties of our solution in Sec.~\ref{sec:scal}. The fluctuating induction equation at first order is
\beq
{\vb}_{0\tau} + \left[ \vpu (\vA+\vb_0)\right]_\lambda + \alpha\hat{\boldsymbol{x}}\vb_{0x} = \vap(\vu_1-\vb_1)_\lambda\label{eq:ind1}.
\eeq
Inserting the zeroth-order solution $\vu_0 = \vb_0$ and taking the sum of (\ref{eq:mom1}) and (\ref{eq:ind1}), the RHS cancels, and we are left with
\begin{align}
2{\vb}_{0\tau} &+ \left[ \vpu (\vA+\vb_0)\right]_\lambda \nonumber\\&+ \vp\left[ c_s^2\frac{\rho_1}{\avg\rho}+ {\vb_1\cdot(\vA+\vb_0)}\right]_\lambda + \alpha\vb_0 = 0.\label{eq:withpress}
\end{align}
Taking the dot-product of this equation with $\vp$ gives a relationship between the first-order compressive fluctuations and the radial component of the zeroth-order magnetic-field fluctuation,
\beq
\left[ c_s^2\frac{\rho_1}{\avg\rho}+ {\vb_1\cdot(\vA+\vb_0)}\right]_\lambda=  \frac{2\alpha p_xb_{0x}}{p^2}  - \frac{(\vpul)(\vap)}{p^2},\label{eq:pmom}
\eeq
where we have used
\beq
\vp \cdot {\vb}_{0\tau} = \left(\cancel{\vp\cdot\vb_0}\right)_\tau- \vb_0\cdot{\vp}_\tau = -\ada p_xb_{0x},
\eeq
using Eq.~(\ref{eq:pb}) to express this in terms of $b_{0x}$ only. This allows us replace the total-pressure gradient term in (\ref{eq:withpress}), giving
\beq
\boxed{{\vb_{0\tau}}+\frac{\alpha}{2}\left(\vb_0+\frac{2\vp p_x}{p^2}b_{0x}\right)+\frac{1}{2}\left[\vp\cdot\vu_1(\vAt+\vb_0)\right]_\lambda = 0,\label{eq:sbeq}}
\eeq
where $\vAt=\vA-\vp(\vap)$ are the components of $\vA$ transverse to $\vp$. This equation encodes the slow evolution of an arbitrary-amplitude Alfv\'en wave in the expanding solar wind. However, this evolution is not yet determined; to proceed, we need to write the compressive component $\vpu$ in terms of $\vb_0$.

\subsection{Finding $\vpu$}
Equation~(\ref{eq:sbeq}) comprises two independent equations for three unknowns (the two transverse components of $\vb_0$ and $\vpu$); so we need one more equation to determine the evolution. This turns out to be provided by the condition (\ref{eq:funny}) that the zeroth-order magnetic field strength is spatially uniform. Let us work in terms of the total zeroth-order magnetic field $\vb=\vA+\vb_0$. Taking the partial derivative of $\vb(\lambda,t,\tau)$ with respect to $\tau$ at constant $\lambda$ and $t$, we obtain
\beq
{\vb}_\tau = {\vb}_{0\tau} - \alpha\hat{\boldsymbol{x}}\vAx.
\eeq
Inserting this into (\ref{eq:sbeq}) and taking the dot product with $\vb$, we obtain
\begin{align}
\frac{1}{2}({b^2})_\tau&+\alpha\left[b_x\vAx+\frac{p_x\vap}{p^2}b_{0x}
+\frac{1}{2}\vb_0\cdot\vb\right]\nonumber\\ &+\frac{1}{2}\vpul|\vAt+\vb_0|^2=0,
\end{align}
where $|\vAt+\vb_0|$, the zeroth-order transverse magnetic field strength, is spatially constant (because $\vp\cdot\vb=\vap$ is too, cf. Eq.~\ref{eq:divb}). We now take the fluctuating part of this equation by subtracting its spatial mean; using the identity
\beq
\fluc{\vb_0\cdot\vb} = \fluc{\vb\cdot\vb}-\fluc{\vA\cdot\vb}=\cancel{\fluc{b^2}}-\cancel{\fluc{v_{\rm A}^2}} - \vb_0\cdot\vA,
\eeq
(the first two crossed-out terms on the RHS are zero because both $b^2$ and $\vA$ are spatial constants) we are left with 
\beq
\boxed{\vpul = - 2\alpha \frac{b_{0x}\left(\vAx + {p_x\vap}/{p^2}\right)-\frac{1}{2}\vb_0\cdot\vA}{|\vAt+\vb_0|^2},\label{eq:u1app}}
\eeq
an equation for the divergence of the first-order velocity in terms of $\vb_0$. Equations (\ref{eq:sbeq}) and (\ref{eq:u1app}) determine the evolution of $\vb_0$ on the expansion timescale,\footnote{Combined, of course, with a suitable initial condition at $\tau=0$, and the appropriate periodic boundary condition in $\lambda$.} and are the main result of this paper.
\subsection{Magnetic pressure fluctuations}
As stated in Eq.~(\ref{eq:nearlyalf}), the magnetic pressure is constant at zeroth order. At first order, the gradient of the magnetic pressure is (in Alfv\'en units) $\epsilon\vp[\vb_1\cdot(\vA+\vb_0)]_\lambda$. Using (\ref{eq:dens}), (\ref{eq:pmom}) and (\ref{eq:u1app}), we may write this solely in terms of $\vb_0$,
\begin{widetext}
\begin{align}
\boxed{[\vb_1\cdot(\vA+\vb_0)]_\lambda = \frac{2\alpha}{p^2} \left[ p_xb_{0x} + \vap\left(1-\frac{\beta}{\cos^2\theta}\right)\frac{b_{0x}\left(\vAx + {p_x\vap}/{p^2}\right)-\frac{1}{2}\vb_0\cdot\vA}{|\vAt+\vb_0|^2}\right],\label{eq:magpressapp}}
\end{align}
\end{widetext}
where $\theta$ is the angle between $\vp$ and $\vA$. This equation describes how expansion creates a perturbation in the magnetic pressure.
\subsection{Physical summary}\label{sec:derivsumm}
We will now briefly summarise our results before analysing their consequences in the next section. We collect the important equations (those which are boxed where they appear in the derivation) here for clarity. Our main result is a dynamical equation (\ref{eq:sbeq}) governing the slow nonlinear evolution of an arbitrary-amplitude Alfv\'enic fluctuation $\vb_0$ as it travels out in the expanding plasma of the solar wind:
\beq
{\vb_{0\tau}}+\frac{\alpha}{2}\left(\vb_0+\frac{2\vp p_x}{p^2}b_{0x}\right)+\frac{1}{2}\left[\vp\cdot\vu_1(\vAt+\vb_0)\right]_\lambda = 0.\label{eq:sbeqmain}
\eeq
The $\tau$-derivative term is the time derivative on the slow expansion timescale. The $\alpha \vb_0 /2$ term gives rise to the slow decay of the wave amplitude on the expansion timescale (see Sec. \ref{sec:ampsc}). The term in the $\vp$-direction encodes the divergence-free nature of the magnetic fluctuations as the direction of the gradients, $\vp$, rotates towards radial due to expansion. The final term is nonlinear, and describes the distortion of the primary Alfv\'enic fluctuation as it increases in normalized amplitude, by a compressive velocity \beq
u_{1\mathrm{comp}} = \epsilon \vpu / p.\label{eq:ucompdef}
\eeq
The gradient of this velocity is also determined as part of our analysis, in Eq.~(\ref{eq:u1app}), which reads
\beq
\vpul = - 2\alpha \frac{b_{0x}\left(\vAx + {p_x\vap}/{p^2}\right)-\frac{1}{2}\vb_0\cdot\vA}{|\vAt+\vb_0|^2}.\label{eq:vpud}
\eeq
This nonlinear distortion steepens the waveform of $\vb_0$, maintaining its constant magnetic field strength $|\vA+\vb_0|$, despite its changing amplitude. Related to this compressive velocity, there are density fluctuations \beq
\delta \rho_1 = \epsilon \rho_1,
\eeq
given using the continuity equation by Eq.~(\ref{eq:dens}), which reads
\beq
\frac{\rho_1}{\avg\rho} = - \frac{\vpu}{\vap},\label{eq:densmain}
\eeq
as well as magnetic-pressure fluctuations 
\beq
\dbsq = 2\epsilon \vb_1\cdot(\vA+\vb_0),
\eeq
determined in terms of their gradient by Eq.~(\ref{eq:magpressapp}), which reads
\begin{widetext}
\begin{align}
[\vb_1\cdot(\vA+\vb_0)]_\lambda = \frac{2\alpha}{p^2} \left[ p_xb_{0x} + \vap\left(1-\frac{\beta}{\cos^2\theta}\right)\frac{b_{0x}\left(\vAx + {p_x\vap}/{p^2}\right)-\frac{1}{2}\vb_0\cdot\vA}{|\vAt+\vb_0|^2}\right].\label{eq:magpress}
\end{align}
\end{widetext}
The physical scenario described by these equations is quite simple: expansion of a nearly-constant-$B^2$ Alfv\'en wave ($\vb_0$) constantly drives a small fluctuation in the magnetic pressure $\dbsq$ (\ref{eq:magpress}); this drives a compressive flow $\ucomp$ (\ref{eq:vpud}) which nonlinearly distorts the main Alfv\'enic fluctuation via (\ref{eq:sbeqmain}), keeping the fluctuations in the magnetic pressure small and thus largely preserving the Alfv\'enic character of the wave.

\section{Scalings}\label{sec:scal}
There are several interesting and simple scaling properties shared by all solutions that can be deduced from Eq.~(\ref{eq:sbeqmain}) and the expressions for the compressive components (\ref{eq:ucompdef}--\ref{eq:magpress}). We examine these in this section.
\subsection{Amplitude scaling}\label{sec:ampsc}
First, notice that
\begin{align}
\left[\vp\cdot\vu_1(\vAt+\vb_0)\right]_\lambda\cdot \vb_0 = &\left[\vp\cdot\vu_1(\vAt+\vb_0)\cdot \vb_0\right]_\lambda \nonumber\\
&-\vp\cdot\vu_1\cancel{(\vAt+\vb_0)\cdot \vb_{0\lambda}},\label{eq:totalderiv}
\end{align}
where the cancellation comes from our assumption (\ref{eq:nearlyalf}) that the zeroth order magnetic-field strength is constant. Taking the dot product of Eq.~(\ref{eq:sbeqmain}) with $\vb_0$ and using (\ref{eq:totalderiv}), we obtain
\beq
\frac{\pd b_0^2}{\pd \tau} + \alpha \left( b_0^2 + \cancel{\frac{2\vp\cdot\vb_0 p_x b_{0x}}{p^2}}\right) + \left[\vp\cdot\vu_1(\vAt+\vb_0)\cdot \vb_0\right]_\lambda = 0,\label{eq:b02}
\eeq
where the cancellation is because $\vp\cdot\vb_0=0$, the expression of $\nabla\cdot \vB = 0$ in this one-dimensional system. Upon performing a spatial average of (\ref{eq:b02}) using (\ref{eq:avgdef}), the nonlinear term vanishes since it is a total derivative, leaving
\beq
\pddtau{\avg{b_0^2}}= -\alpha \avg{b_0^2},\label{eq:ampsc}
\eeq
whose solution is 
\beq
\avg{b_0^2}\propto a^{-1}.\label{eq:ampsc2}
\eeq
This dependence is identical to the WKB solution at low amplitude \citep{grappin1993}. The only difference is that now it is the \emph{average} amplitude that has the scaling law: this is because the waveform must distort in order to keep $B^2$ nearly constant. Then, using (\ref{eq:ampsc2}) and (\ref{eq:meanva}), the normalized RMS amplitude, in the case of radial $\avg{\vB}$, scales as
\beq
\sqrt{\frac{\avg{b_0^2}}{v_{\rm A}^2}} \propto a^{1/2},\label{eq:ampsc3}
\eeq
i.e., the normalized RMS amplitude increases with $a$. 
This is a special case (with the solar wind velocity $U$ constant) of the result previously obtained by \citealt{hollweg1974} (in a more general case with the solar wind velocity $U\ne \text{const.}$ and super-radial expansion of the magnetic field lines) and \citealt{barnes1974} (in a much more general case, without any symmetry). However, these authors did not study explicitly how the wave is distorted by the expansion. 

What does this mean (or more generally, what does the \citealt{hollweg1974} result mean) for radial magnetic-field reversals (an important aspect of the observed switchbacks)? To reverse the field, the fluctuation must grow to high amplitude. Let us assume that the source of the Alfv\'en waves at the coronal base is statistically quite uniform, with a low initial amplitude. Then there are two rather obvious consequences. 

First, sufficiently close to the Sun, we should expect the fraction of the volume filled with switchback fluctuations to be an increasing function of $R$:\footnote{Probably within a few tens of $R_\odot$. This is for two reasons; first, further from the Sun the non-radial components of the mean field become more important (the ``Parker Spiral"). These decay more weakly with $a$ (cf. Eq.~\ref{eq:meanva}), and so the normalised amplitude tends to saturate. Second, further from the Sun, our neglect of nonlinear interaction due to counterpropagating Alfv\'en waves produced by reflection (i.e. turbulence), as well as parametric decay and/or other instabilities, becomes less justifiable: these would also cause a more significant decay in the fluctuation amplitude.} the first switchbacks to form are the fluctuations with initially largest amplitude, and as the expansion continues more and more fluctuations can form switchbacks. This is indeed what is seen in the observations \citep{mozer2020,badman2020b}, which seems to be strong evidence in support of the \emph{in situ} generation hypothesis \citep{squire2020}: it is hard to imagine a similar dependence emerging from the mechanisms in which the switchbacks have a lower-coronal origin \citep{drake2020,zank2020}. 

Second, going beyond our Eq.~(\ref{eq:ampsc3}) and applying the \citet{hollweg1974} results, we should expect more reversals in regions where larger super-radial expansion of the magnetic field occurs. This reduces $\avg{B}$, which thus acts to increase $\delta B / \avg{B}$. Thus, switchbacks should be observed in ``patches" that are correlated with regions where the expansion has been most significant (e.g. slow wind), and they should be relatively absent in regions where the expansion has not been as dramatic (e.g. fast wind). This is a specific, falsifiable prediction of the \citet{squire2020} Alfv\'en wave expansion model for switchback generation.

%It should be noted that the amplitude evolution in (\ref{eq:ampsc3}) is not particularly novel (since, as we have explained, a more general case has been known since 1974), and could be regarded as merely a check that we are on the right track. However, the interpretation advanced here in terms of its relevance to the magnetic-field reversals and switchbacks observed by PSP, may be of some interest. The later parts of our analysis, regarding the distortion and compressive components of the wave, are (to our knowledge) new.

\subsection{Parallel magnetic field fluctuations}\label{sec:b0xsc}
First, we define an additional coordinate system. Let us define unit vectors $\vn = \vp \times \vA/pv_A$ and $\boldsymbol{m}=(\vp\times\vA)\times\vp/p^2v_A$; the mean field $\vA$ is in the $\boldsymbol{m}$-$\vp$ plane, and $\vAm=v_{\rm A}\sin\theta$ and $v_{\mathrm{A}p} = v_{\rm A} \cos\theta$, where $\theta$ is the angle between $\vp$ and $\vA$. Then, the constancy of the zeroth-order transverse (to $\vp$) magnetic field strength, $\btr^2=|\vAt+\vb_0|^2=\text{const.}$, means that magnetic field fluctuations in the $m$-direction are related to $\btr$ and the fluctuations in the $n$-direction by%\footnote{Two simple (but pathological) solutions are, first, $b_{0n}=\pm\sqrt{\btr^2-v_{Am}^2}$, $b_{0m}=0$, and second, $b_{0m}=-\vAm\pm \btr$, $b_{0n}=0$, with the sign in both cases flipping as a function of $\lambda$ to make the average over the whole period $\Lambda$ zero. We exclude these completely discontinuous, square-wave solutions; they are not produced by the evolution of any of our initial conditions.}
\begin{align}
\btr^2 &= b_{0n}^2 + (\vAm+b_{0m})^2, \nonumber\\
&=b_{0n}^2 + 2\vAm b_{0m} + \vAm^2+b_{0m}^2.\label{eq:bz}
\end{align}
Let us also define the normalized amplitude,\footnote{Note that this is normalized to the mean field $v_{\rm A}$, rather than the total field strength $b$: $A$ is unbounded, whereas $b_0/b\leq 2$ for a system with $b$ constant \citep{matteini2018}.}
\beq
A= \frac{1}{v_{\rm A}}\sqrt{{b_{0n}^2}+{b_{0m}^2}},\label{eq:Adef}
\eeq
and also
\beq
D=\frac{A}{\sin\theta}.\label{eq:Ddef}
\eeq
From Eq.~\ref{eq:Adef}, we trivially have
\beq
\frac{b_{0m}}{v_{\rm A}} \lesssim A.\label{eq:b0mhighamp}
\eeq
However, for $D\ll1$, a more restrictive bound applies, which may be derived as follows. In terms of $D$, Eq.~(\ref{eq:bz}) may be written
\beq
\frac{\btr^2}{\vAm^2} = D^2 + 1 + \frac{2 b_{0m}}{\vAm}.\label{eq:bz2}
\eeq
Taking the spatial average of this equation (cf. Eq.~\ref{eq:avgdef}), 
\beq
\frac{\btr^2}{\vAm^2} = \avg{D^2} + 1,\label{eq:avbz2}
\eeq
and so
\beq
\frac{b_{0m}}{\vAm}=\frac{1}{2}\left(\avg{D^2}-D^2\right).\label{eq:bmDsc}
\eeq
%Making use of the fact that the linearized Alfv\'en wave is polarized in the $\boldsymbol{n}$ direction, we assume from now on that $b_{0n}\gtrsim b_{0m}$. 
Now, let us look at small-amplitude case specifically, $D\ll1$. If we assume that in terms of scalings, $D^2 \sim \avg{D^2}$, we see that $b_{0m}/\vAm$ is second-order in $D$ or smaller (and upon inspecting Eq.~(\ref{eq:Adef}), $b_{0n}/\vAm$ is first-order in $D$). Using (\ref{eq:Adef}), (\ref{eq:Ddef}), (\ref{eq:bmDsc}), and $\vAm=v_{\rm A}\sin\theta$, in terms of scaling with $A$ we have
\beq
\frac{b_{0m}}{v_{\rm A}} \lesssim \frac{A^2}{\sin\theta}, \quad \text{when }D\ll1.\label{eq:b0mlowamp}
\eeq
This relationship is well-known \citep{barnes1974,vasquez1998b}. As $D\to0$, the polarization of the linearized Alfv\'en wave is recovered. As $D\to1$ from below, (\ref{eq:b0mlowamp}) matches the less restrictive earlier bound (\ref{eq:b0mhighamp}); for $D\gtrsim1$, $b_{0m}/v_{\rm A}$ is no longer required to be much smaller than $b_{0n}$, and must merely satisfy (\ref{eq:b0mhighamp}). Both our scalings (\ref{eq:b0mlowamp}) and (\ref{eq:b0mhighamp}) put upper bounds on the scaling of $b_{0m}$; the exact relationship between $b_{0m}$ and $b_{0n}$ depends on the waveform. We proceed by assuming that $b_{0m}$ agrees with the scaling corresponding to smallest applicable upper bound\footnote{This will be true in the numerical simulations in Sec.~\ref{sec:evol}, and is true for all simulations we have tried starting with small-amplitude, continuous initial conditions.}.

The component of the fluctuation parallel to the magnetic field $\avg{\vB}$ is
\beq
b_{0\parallel} = b_{0m}\sin\theta,
\eeq
because the angle between $\vp$ and $\vA$ is $\theta$, and $\boldsymbol{m}$ is perpendicular to $\vp$ (the wave is transverse) but in the plane of $\vp$ and $\vA$. Using the scalings of $b_{0m}$ described above (Eqs.~\ref{eq:b0mlowamp} and \ref{eq:b0mhighamp}),
\beq
\frac{b_{0\parallel}}{v_{\rm A}} \sim \begin{cases}
A^2, & D\ll 1\\
A\sin\theta, & D\gtrsim1,\label{eq:b0par}
\end{cases}
\eeq
which matches at $D\sim 1$. In this paper, we will only consider radial $\vA$, and so this component of the fluctuations is responsible for magnetic-field reversals. In the following section, we will make use of these scalings to examine the scaling behaviour of the compressive components. Moreover, (\ref{eq:b0par}) shows that there is an angular dependence to the propensity of fluctuations to reverse the background field: for a switchback, $b_{0\parallel}/v_A \gtrsim 1$, which is attained at lower amplitude $A$ for perpendicular ($\sin\theta\approx 1$) waves. This may help to explain the observations of \citet{laker2020}, who found that switchbacks have aspect ratios of around $10$--$30$, extremely elongated in the direction of the mean magnetic field.

\subsection{Compressive velocity and density fluctuations}\label{sec:vpusc}
Assuming a radial $\vA=\hat{\boldsymbol{x}}\vAx$, we may write (\ref{eq:vpud}) as
\beq
p\ucompl = -2\epsilon\alpha \frac{b_{0x}\vAx(\cos^2\theta+1/2)}{b_T^2}.
\eeq
Recall that $\alpha/\vap=1$ (Eq.~\ref{eq:epsconst}); then,
\beq
\ucompl = -\epsilon\left(2\cos^2\theta+1\right)\cos\theta \left( \frac{\vAx^2}{b_T^2}\right)b_{0x}.\label{eq:ucompeps}
\eeq
We can estimate the amplitude of the compressive velocity compared to that of the Alfv\'enic velocity fluctuations $\vu_0=\vb_0$. First, note that if the characteristic width of the waveform in $\lambda$ is roughly constant, $\ucompl\sim \ucomp$. Then at low amplitude, $u_0\ll \vAx\sin\theta$, using (\ref{eq:b0par}),
\beq
\frac{\ucomp}{u_0} \sim \epsilon\left(2\cos^2\theta+1\right)\frac{\cos\theta}{\sin^2\theta} \frac{u_0}{\vAx},\label{eq:uclowamp}
\eeq
At high amplitude, $u_0\gg\vAx\sin\theta$, we instead obtain
\beq
\frac{\ucomp}{u_0} \sim \epsilon\left(2\cos^2\theta+1\right)\sin\theta\cos\theta\frac{\vAx^2}{u_0^2}.\label{eq:uchighamp}
\eeq
First, note that the expressions match at $u_0\sim\vAx\sin\theta$ (i.e. $D\sim1$), as expected. 
Second, we see that the compressive velocity scales with $\epsilon$, or equivalently $\dot{a}$: thus, we might expect more compressive fluctuations in regions which undergo larger expansion. Interestingly, the slow solar wind, which has expanded more, is observed to have a larger compressible fraction \citep{bruno2013} than the fast wind, which is thought to have expanded less \citep{wang1990,chandran2021}. 
\begin{figure*}
\includegraphics[width=\linewidth, trim=0 0.5cm 0 1.5cm, clip]{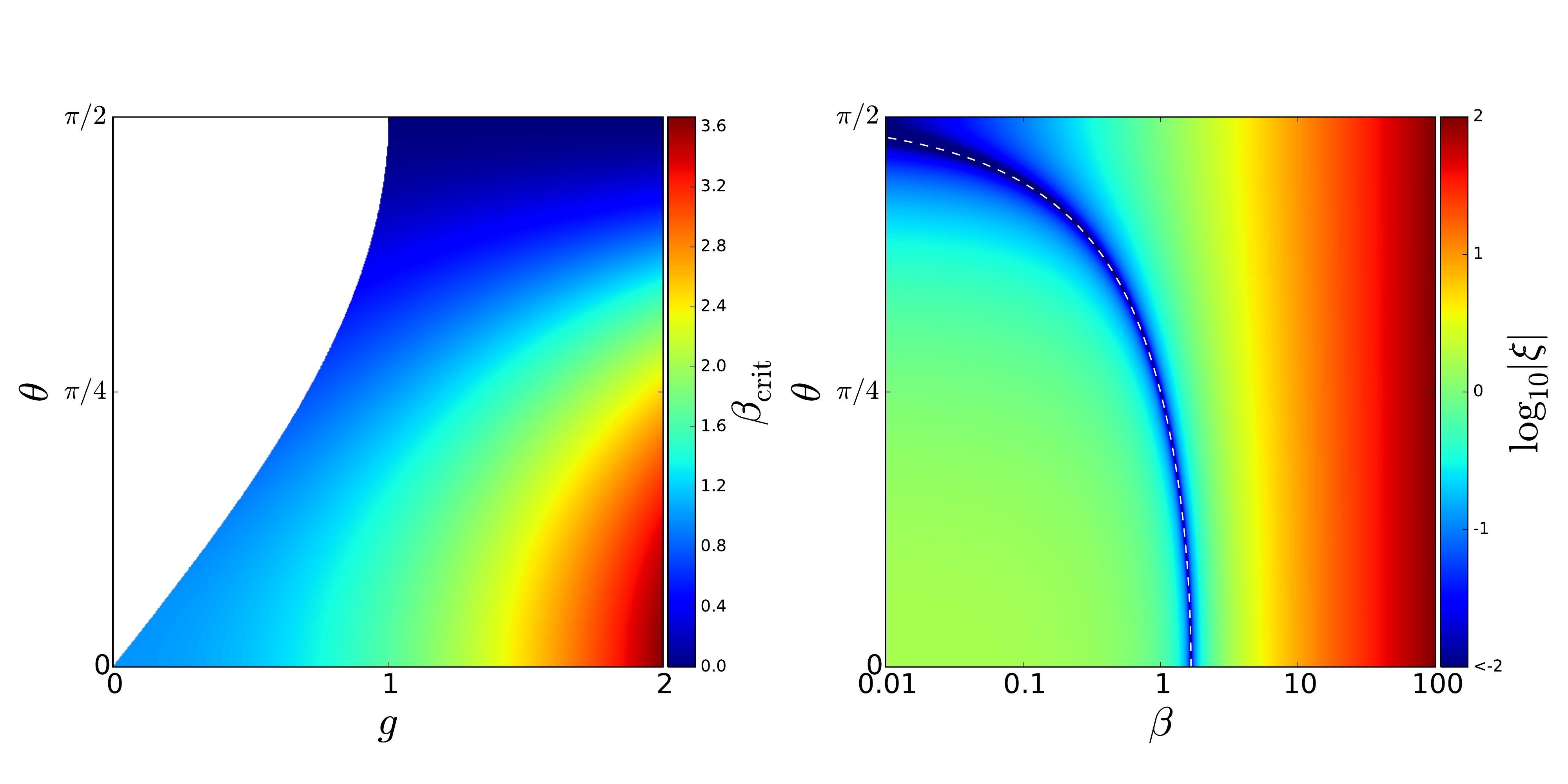}
\vskip-0.2cm
\caption{\textbf{Left:} $\beta_{\rm crit}$ plotted as a function of $g=b_T/\vAx$ and $\theta$. Note that $g\geq\sin\theta$ by geometry; accordingly, the forbidden region is left unfilled on the plot. \textbf{Right:} The logarithm of $|\xi|$ (Eq.~\ref{eq:chi}) plotted as a function of $\beta$ and $\theta$, for $g=1$; to the left of the white dotted line (which is centered around $\beta=\beta_{\rm crit}$), $\xi>0$, while to the right, $\xi<0$.\label{fig:bcrit}}
\end{figure*}

Let us assume that the initial propagation angle $\theta^*$ at the coronal base where the wave originates is nearly perpendicular, $0<\cos\theta^*\ll1$. Then, initially $\cos\theta\sim 1/\tan\theta \sim a/\tan\theta^*\ll1$ (using Eq.~\ref{eq:prot}). Eventually, $\theta$ decreases until $\cos\theta\approx 1$, and instead $\sin\theta\propto 1/a$. Thus, also using $u_0/\vAx \propto a^{1/2}$ (cf. Eq.~\ref{eq:ampsc}), we expect
\beq
\frac{\ucomp}{u_0} \propto \begin{cases}
a^{3/2}, \quad D\ll1\\
a^{-2},\quad D\gg1.
\end{cases}\label{eq:ucompa}
\eeq
The maximum is therefore reached at approximately $D\sim1$; thus, $A\sim\sin\theta$ is the characteristic amplitude at which the Alfv\'en waves are maximally steepened and distorted by the expansion in order to keep $B^2$ constant.

Via the continuity equation, the density fluctuations have a rather simple relationship with the compressible component of the velocity, given by Eq.~(\ref{eq:densmain}),
\beq
\frac{\delta\rho_1}{\avg\rho} = -\frac{\ucomp}{\vAx\cos\theta}.\label{eq:rhoucomp}
\eeq
Thus, for rather perpendicular waves for which $\cos\theta\ll1$ (as appears to be the case for switchbacks in the real world, cf. \citealt{laker2020}), the density fluctuations should be large compared to the compressive velocity. Given an observational estimate for $\theta$ (for example, using a minimum variance analysis), this is a testable prediction of our model.

\subsection{Magnetic field strength fluctuations}\label{sec:b2dsc}
The fluctuations in magnetic field strength are given by (\ref{eq:magpress}). With radial $\vA$, this reduces to
\beq
\frac{1}{2}\delta (b^2)_{1\lambda} = 2\epsilon\alpha\frac{p_xb_{0x}}{p^2} \frac{2b_T^2+\left(1-\frac{\beta}{\cos^2\theta}\right)\vAx^2(2\cos^2\theta+1)}{2b_T^2}.
\eeq
Using (\ref{eq:epsconst}), we find 
\beq
\frac{\delta (b^2)_{1\lambda} }{\vAx^2}= \epsilon \frac{b_{0x}}{\vAx} \frac{2b_T^2\cos^2\theta+\left(\cos^2\theta-\beta\right)\vAx^2(2\cos^2\theta+1)}{b_T^2}.\label{eq:magpresseps}
\eeq
The transverse magnetic field $b_T$ is spatially constant; let us set it equal to $g\vAx$, where $g$ is a function of $a$ that depends on the specific solution. Then, the magnetic pressure fluctuation vanishes if 
\begin{align}
\beta=\beta_{\rm crit}=\cos^2\theta + \frac{2g^2\cos^2\theta}{2\cos^2\theta+1}.
\end{align}
As $\theta\to0$, $\beta_{\rm crit}\to1+2g^2/3$, while as $\theta\to\pi/2$, $\beta_{\rm crit}\to0$. For $b_0\sim\vAx$, $g$ is of order unity; thus, the magnetic pressure fluctuations in large-amplitude waves of moderate to parallel $\theta$ are always minimized for $\beta$ of order unity. For clarity, $\beta_{\rm crit}$ is plotted as a function of $g$ and $\theta$ in the left-hand panel of Figure~\ref{fig:bcrit}; we indeed see that $\beta_{\rm crit} \sim 1$ apart from close to $\theta=\pi/2$. This explains the numerical observation of \citet{squire2020}, who saw that the magnetic compression $C_{B^2}=\langle\delta(B^2)\rangle/\langle|\delta \vB|^2\rangle$ was indeed minimized in their expanding box turbulence simulations when $\beta \sim 1$. Observationally, there are hints of a similar effect in \citet{larosa2020}, as we discuss in the next section. 

\subsection{Compressive component polarization}\label{sec:polsc}
\citet{larosa2020} have performed a statistical analysis of 70 switchbacks, and found that in some the density and field-strength fluctuations have the same sign, but in others they have opposite sign. This may be a natural consequence of the $\beta$-dependence identified in the previous section. Use Eqs.~(\ref{eq:ucompeps}), (\ref{eq:rhoucomp}) and (\ref{eq:magpresseps}) to form the polarization ratio
\begin{align}
\xi &= \frac{\dbsqm/\vAx^2}{\rho_{1\lambda}/\avg{\rho}} \nonumber\\&= \frac{2({b_T^2}/{\vAx^2})\cos^2\theta+\left(\cos^2\theta-\beta\right)(2\cos^2\theta+1)}{2\cos^2\theta+1}.\label{eq:chi}
\end{align}

For $\beta<\beta_{\rm crit}$, $\xi>0$, while for $\beta>\beta_{\rm crit}$, $\xi<0$. This is plotted in the right-hand panel of Figure~\ref{fig:bcrit}: we plot $\log_{10}|\xi|$ as a function of $\beta$ and $\theta$, choosing $g=1$ (appropriate for moderately large-amplitude waves). The blue band around $\beta_{\rm crit}$ (shown as a dashed white line) is the region where the magnetic-field-strength fluctuations are very low; to the left of this band (low $\beta$), $\xi>0$ and tends to a constant as $\beta\to0$; while to the right (high $\beta$), $\xi<0$ and is unbounded ($\xi\propto\beta$) as $\beta\to\infty$. Thus, it is potentially the combination of $\beta$ and propagation angle that causes the highly variable behaviour in the observations. It should be possible to diagnose $\theta$ in the data using a minimum variance analysis, which would make a test of this hypothesis possible. Given that we have restricted ourselves to a locally isothermal equation of state, it is unlikely that (\ref{eq:chi}) will be reproduced exactly. For example, \citet{farrell2020} found opposite trends in pressure and density across switchback boundaries, which is impossible with our isothermal model. However, the basic principle involved, that $\beta$ and $\theta$ will affect the polarization of the compressive components of the wave, should remain valid.

This transition also appears in the early work of \citet{cohen1974}, who studied the relaxation of quasi-parallel, low-amplitude waves to constant-$B^2$ Alfv\'enic states. In their theory, it appears as a singularity: at $\beta=1$, their evolution becomes infinitely fast (violating their assumptions) because their fluctuations in magnetic field strength are imposed as an initial condition only. In our case, the effect is much less harmful because $\dbsq$ is generated only by the expansion: physically, the interpretation is that around $\beta=\beta_{\rm crit}$, the wave is able to remove $\dbsq$ very efficiently.
\begin{figure*}
\includegraphics[width=\linewidth,trim={0 0.5cm 0 0.2cm},clip]{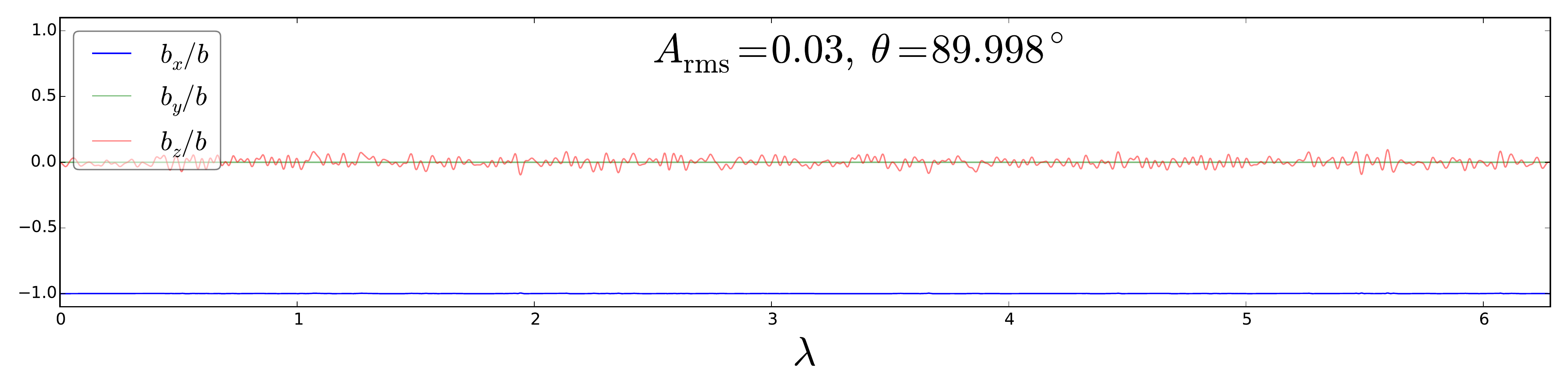}
\includegraphics[width=\linewidth,trim={0 0.5cm 0 0.2cm},clip]{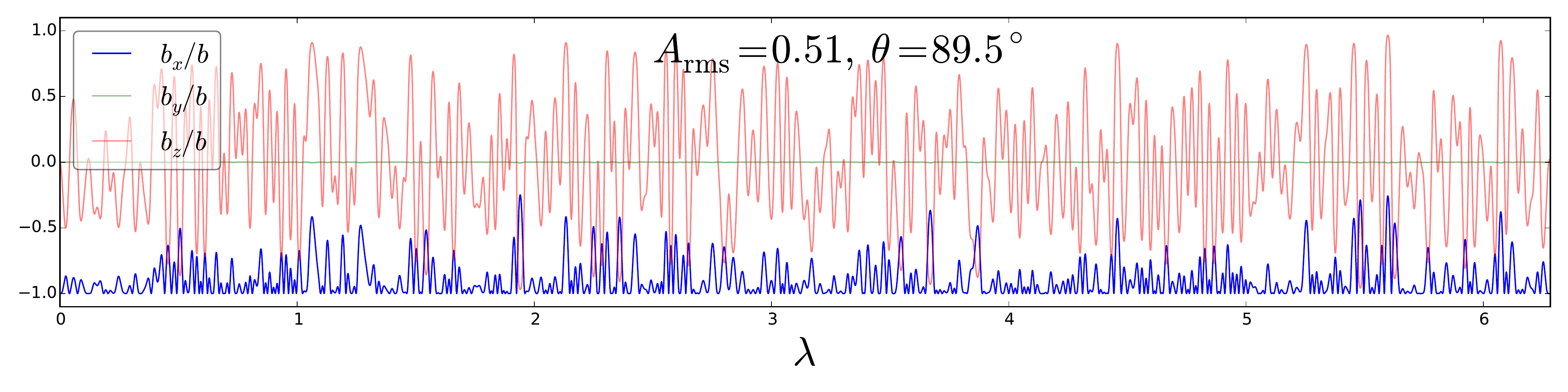}
\includegraphics[width=\linewidth,trim={0 0.5cm 0 0.2cm},clip]{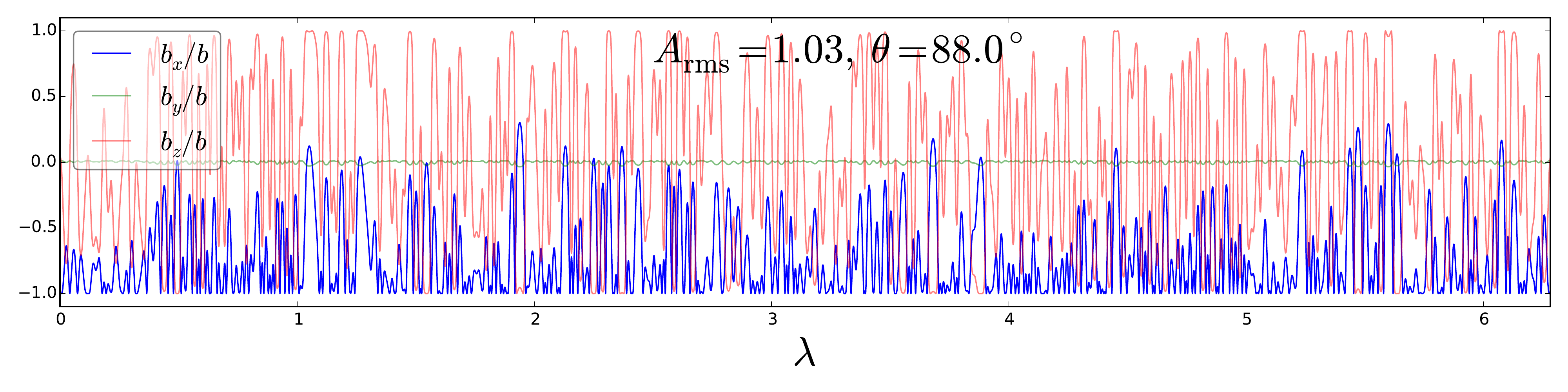}
\includegraphics[width=\linewidth,trim={0 0.5cm 0 0.2cm},clip]{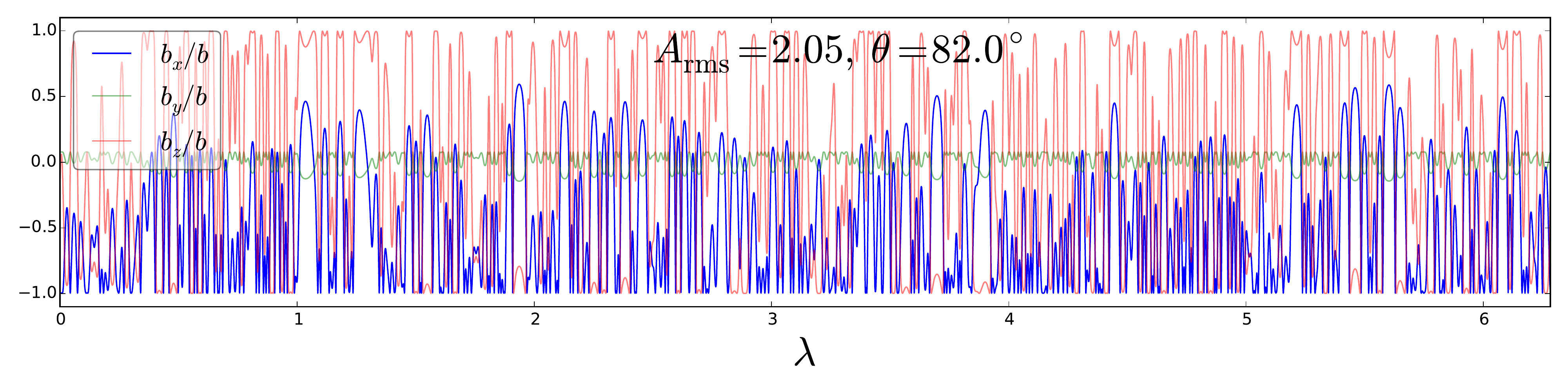}
\caption{Evolution of white noise with initial angle determined by (\ref{eq:thetastartarget}) using parameters (\ref{eq:targetmatchobs}) chosen to approximately match observations, shown at four different times at which the RMS amplitude is $A_{\rm rms} = 0.03,0.51,1.03,2.05$, top to bottom. The angle $\theta$ between $\vp$ and $\vA$ is also shown for each time. The $x$, $y$, and $z$ components of the primary Alfv\'en wave $\vb_0$ are shown in blue, green and red respectively, normalized by the total (zeroth-order) magnetic field strength $b$. \label{fig:matchedobs}}
\end{figure*}
\section{Example solutions}\label{sec:evol}
To illustrate the properties of the wave more concretely, we solve (\ref{eq:sbeqmain}) numerically. We use a pseudospectral method with $n_\lambda = 16384$ grid points, and $\Lambda=2\pi$, with an initial condition $\vb_0^*$ at $\tau=0$, and the compressive components calculated according to Eqs.~(\ref{eq:ucompdef}--\ref{eq:magpress}). We take $\vp$ in the $x$-$y$ plane. We impose white noise in $b_{0n}^*$ in the $\vn = \vp \times \vA/pv_A$ direction, low-pass filtered to keep only the first 200 harmonics. To this, we add another component of the magnetic field in the $\boldsymbol{m}=(\vp\times\vA)\times\vp/p^2v_A$ direction to make this initial condition into a constant-$B^2$ Alfv\'en wave:
\beq
b_{0m}^* = -\vAx^*\sin\theta^* + \sqrt{C^2-b_{0n}^{*2}},\label{eq:b0mstar}
\eeq
so that the initial transverse magnetic field strength is a constant in $\lambda$, $C$. We must choose $C$ so that $\langle b_{0m}^*\rangle = 0$; this can be achieved by iteration, provided that the initial $D^*$ (cf. Eq.~\ref{eq:Ddef}) is sufficiently small \citep{barnes1974}. This initial condition mimics the random, volume-filling Alfv\'en waves that are present in the solar wind (bearing in mind we are limited to one dimensional MHD waves, and additionally neglect turbulence and instability).

There are various parameter choices which apply to all simulations presented here. We take the mean magnetic field to be radial, $\avg{\vB} = \hat{\boldsymbol{x}}\avg{B_x}$. The normalized amplitude (cf. Eq.~\ref{eq:Adef}) of the initial waveform is chosen to be small, with root-mean-square (RMS) value $A_{\rm rms}^*=0.03$. %, comparable to the conditions at the coronal base \citep{depontieu2007,ballegooijen2016}; however, it must be noted that our solutions are not really applicable below the Alfv\'en critical point at $R=R_A$, where the solar wind is accelerating ($U\neq \text{const.}$), and so it is not really meaningful to compare our solutions to reality for $R\lesssim R_A$. 
We choose $\epsilon=0.1$, corresponding roughly to the observed switchbacks at the orbit of PSP \citep{laker2020}.

We must also choose the initial angle between $\vp$ and $\vA$, $\theta^*$. As the evolution progresses, $\theta$ evolves as (cf. \ref{eq:prot}, and remembering that we impose radial $\vA$)
\beq
\tan\theta=\frac{\tan\theta^*}{a},\label{eq:thetasc}
\eeq
so that the wave evolves into more parallel propagation with time. To achieve a particular $\theta_{\rm target}$ when the wave attains a particular rms amplitude $A_{\rm target}$, we use Eqs.~(\ref{eq:ampsc}) and (\ref{eq:thetasc}), and choose
\beq
\theta^* = \arctan\left[\left(\frac{A_{\rm target}}{A_{\rm rms}^*}\right)^2\tan\theta_{\rm target}\right].\label{eq:thetastartarget}
\eeq
\subsection{Comparison with observations}
\begin{figure}
\includegraphics[width=\linewidth]{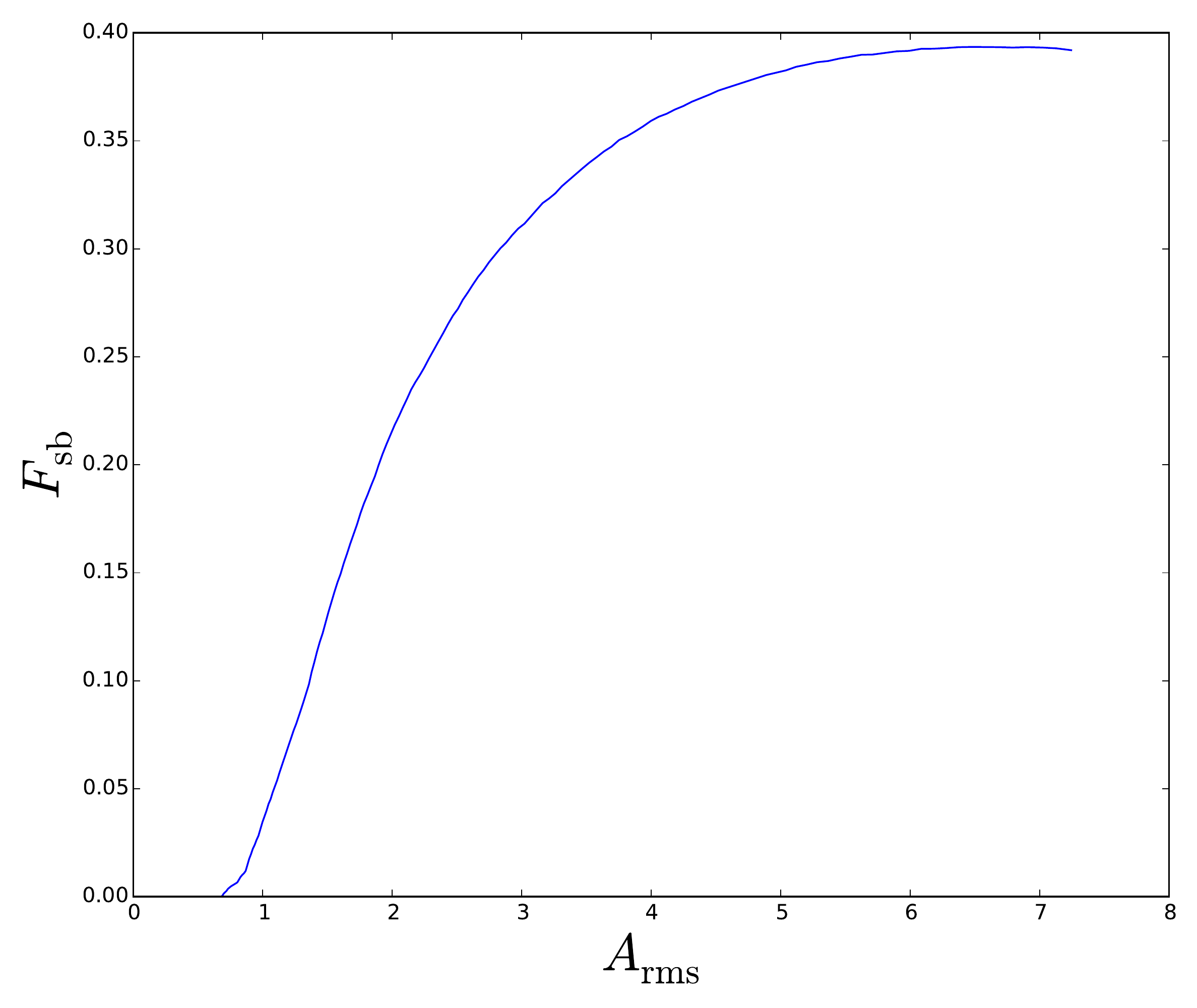}
\caption{Evolution of the switchback fraction $F_{\rm sb}$ with amplitude, for the solution shown in Fig.~\ref{fig:matchedobs}.\label{fig:ampfsb}}
\end{figure}
First, we perform a simulation intended to roughly match the \citet{laker2020} observations, which found that the observed aspect ratio of switchback structures at the orbit of PSP was around $30$. Thus, we choose 
\beq
\tan\theta_{\rm target} = 30, \quad A_{\rm target}=1,\label{eq:targetmatchobs}
\eeq
so that the wave is large-amplitude at $\theta_{\rm target}$.\footnote{$A=1$ is probably a lower bound on the observed fluctuation amplitude at the first few PSP perihelia.} Using Eq.~(\ref{eq:thetastartarget}), this means that $\tan\theta^*$ is extremely large. This probably implies that there must be additional physics involved in the observed large aspect ratio of switchbacks observed by \citet{laker2020}, beyond the simple one-dimensional model presented here (note also that the initial condition is not supposed to correspond accurately to conditions at the coronal base; the EBM is an extremely poor approximation there). However, the solution to our equations that attains $A\sim1$ at $\tan\theta_{\rm target}=30$ may provide some guidance as to the properties of the observed spherically-polarized switchbacks at the orbit of PSP. This solution is plotted at different $A_{\rm rms}$ in Figure \ref{fig:matchedobs}. In its initial state, the wave has only a very small radial field fluctuation (cf. Eq.~\ref{eq:b0mlowamp}); however, once it attains a larger normalized amplitude, the radial fluctuations become relatively large, at some locations reversing the direction of the radial field and creating switchbacks. Superficially, the waveform looks somewhat similar to PSP data (see, for example, Fig.~1 of \citealt{dudokdewit2020}). 

To make this comparison more quantitative, we show in Figure~\ref{fig:ampfsb} the fraction of reversed radial magnetic field (a.k.a. ``switchback fraction", $F_{\rm sb}$) as a function of the normalized amplitude for this simulation. For $A\approx1$, $F_{\rm sb} \approx 0.04$, while for $A\approx 1.5$, $F_{\rm sb} \approx 0.13$: those amplitudes are compatible with the amplitude of the fluctuations at PSP. Thus, the observed $6\%$ of the volume filled by switchbacks at the orbit of PSP \citep{badman2020b} appears to be compatible with simply a collection of randomly phased, large-amplitude AW and does not need any particular special event in the corona. Moreover, the amplitude increases with radial distance from the Sun (cf. Eqs.~\ref{eq:ampsc} and \citealt{hollweg1974}), which then means that the switchback fraction does also: this is also observed \citep{badman2020b,mozer2020}, and would be hard to explain if the structures already reversed the radial field at the point they were generated in the corona \citep{drake2020,zank2020}. However, there is another parameter that is not taken into account in Fig.~\ref{fig:ampfsb}, the angle $\theta$. We examine the effect this has on the propensity to form switchbacks next.
\begin{figure*}
\includegraphics[width=\linewidth,trim={0 0.5cm 0 0.2cm},clip]{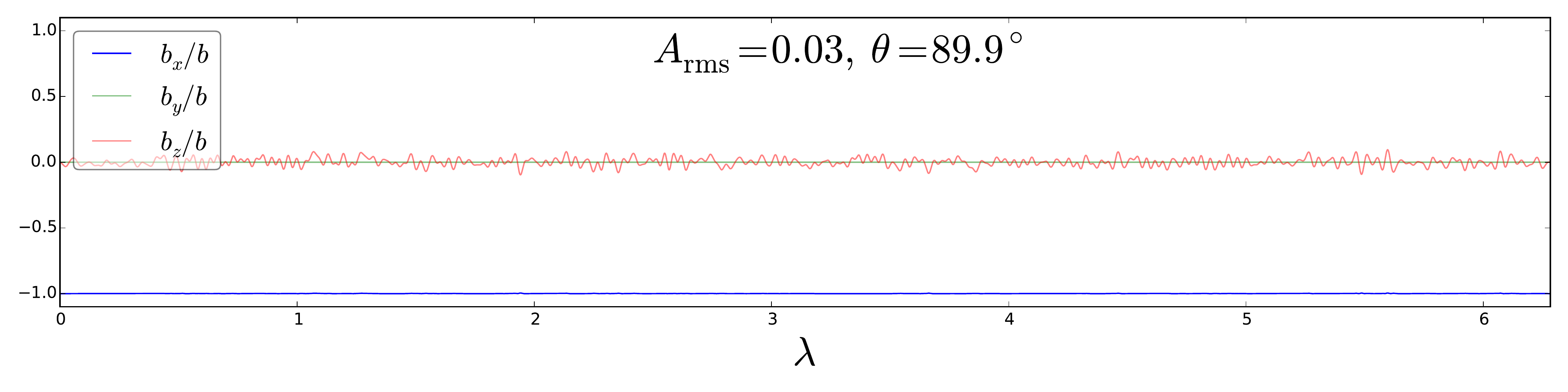}
\includegraphics[width=\linewidth,trim={0 0.5cm 0 0.2cm},clip]{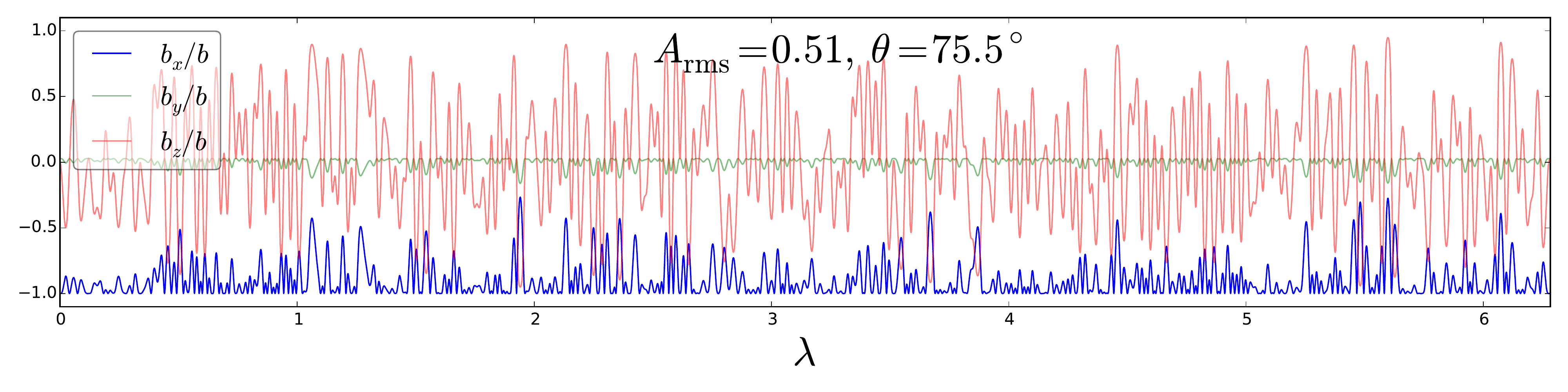}
\includegraphics[width=\linewidth,trim={0 0.5cm 0 0.2cm},clip]{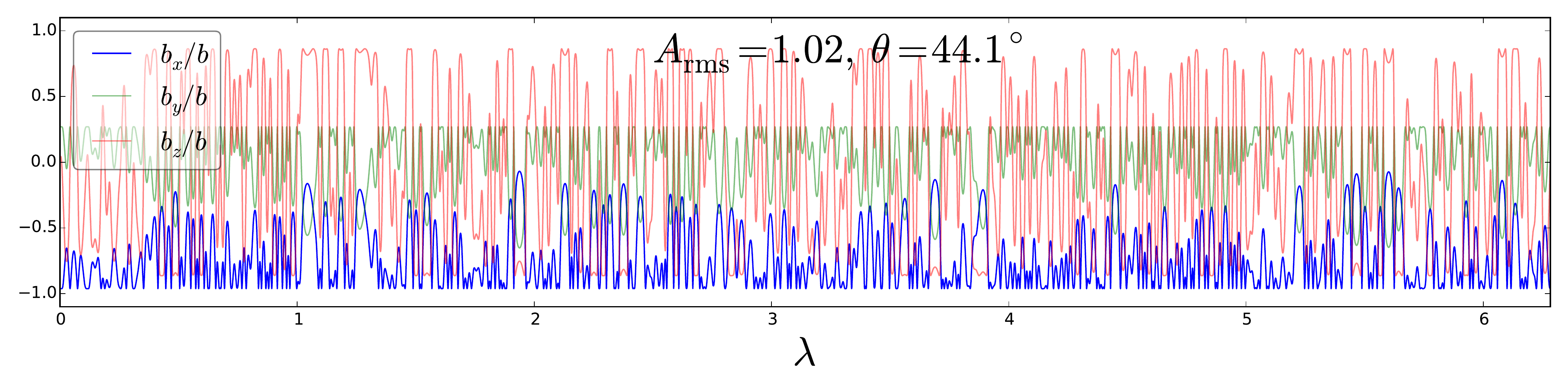}
\includegraphics[width=\linewidth,trim={0 0.5cm 0 0.2cm},clip]{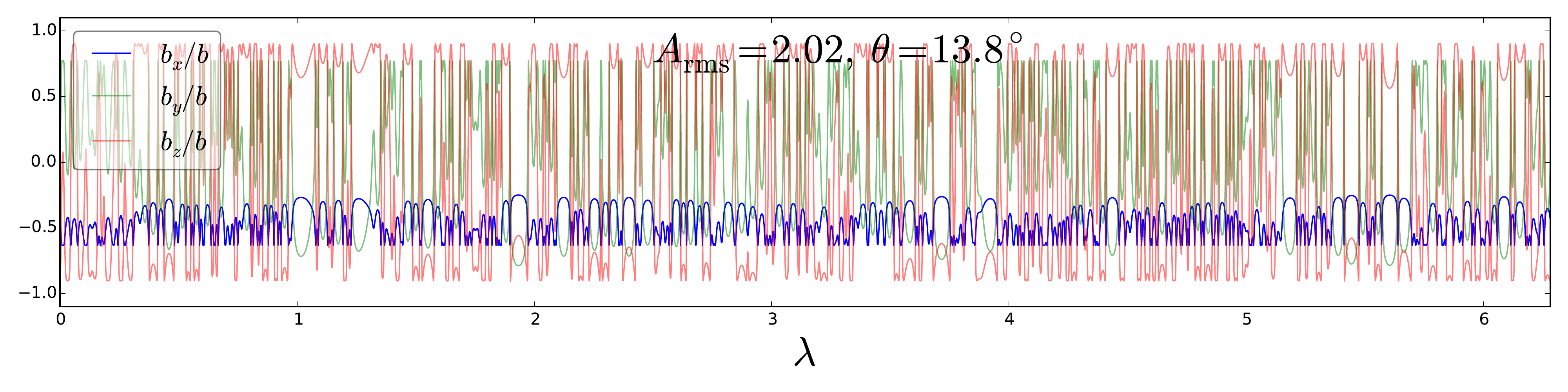}
\caption{Evolution of white noise with initial angle determined by (\ref{eq:thetastartarget}) using parameters (\ref{eq:targetmatchobs}) chosen to approximately match observations, shown at four different times at which the RMS normalized amplitude is $A_{\rm rms} = 0.03,0.51,1.03,2.05$, top to bottom. The angle $\theta$ between $\vp$ and $\vA$ is also shown for each time. The $x$, $y$, and $z$ components of the primary Alfv\'en wave $\vb_0$ are shown in blue, green and red respectively, normalized by the total (zeroth-order) magnetic field strength $b$. \label{fig:lessperp}}
\end{figure*}
\subsection{Effect of $\theta$}
\begin{figure}
\includegraphics[width=\linewidth]{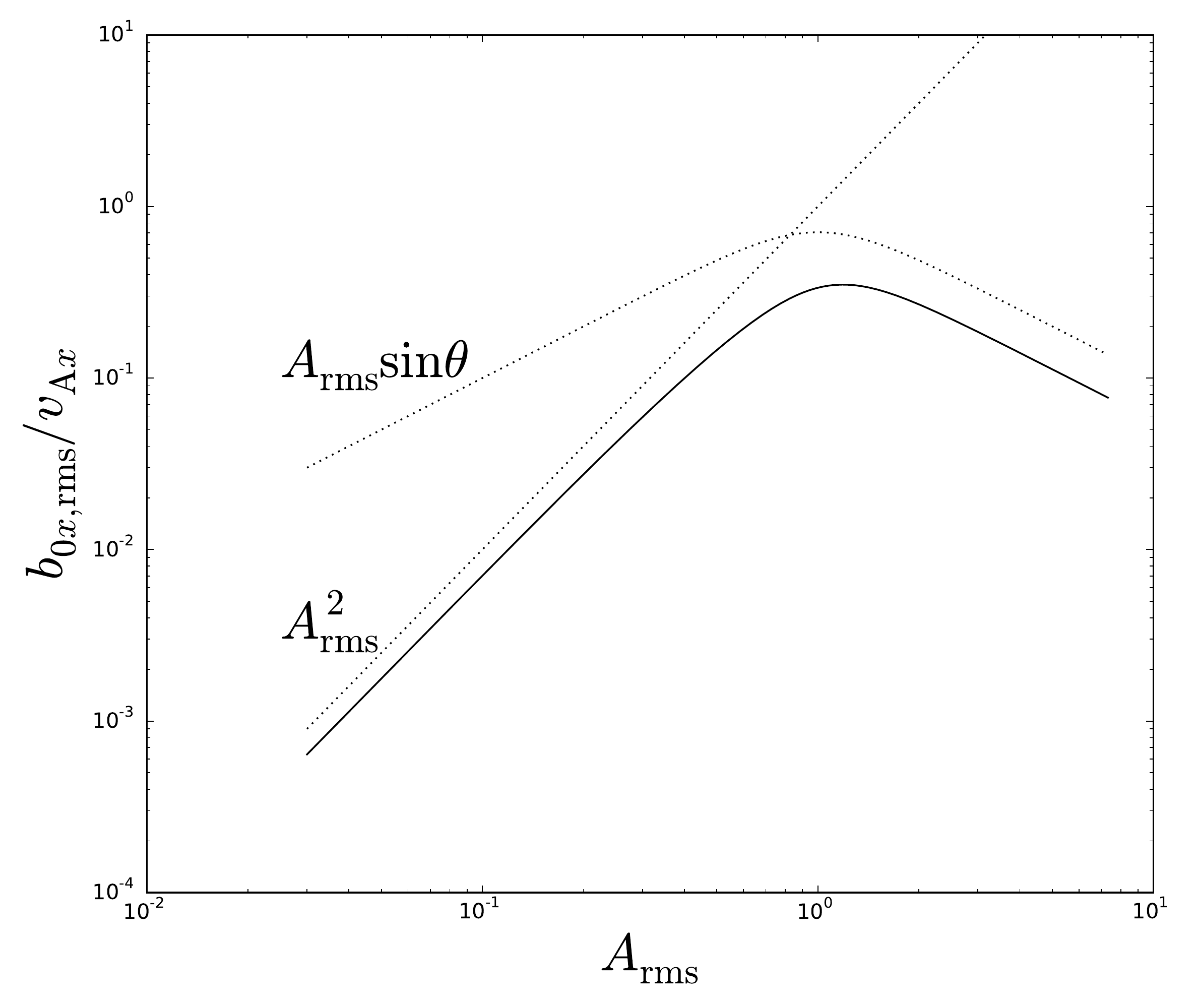}
\caption{The parallel magnetic-field fluctuation $b_{0x,\mathrm{rms}}/\vAx$ (solid line), for the simulation shown in Fig.~\ref{fig:lessperp}. Dotted lines show the two scaling predictions from Eq.~(\ref{eq:b0par}). \label{fig:bxrms}}
\end{figure}

For the simulation shown in Fig.~\ref{fig:matchedobs}, $\theta$ remains close to $90^\circ$ until very late times (large amplitudes). It is very interesting to look at what happens for structures that have a less extreme $\theta^*$, and thus attain a smaller $\theta$ when $A_{\rm rms} =1$. For example, in Fig.~\ref{fig:lessperp} we show results from a simulation with
\beq
\tan\theta_{\rm target} = 1, \quad A_{\rm target}=1,\label{eq:targetlessperp}
\eeq
but the same $b_{0n}^*$ as before: i.e., much less oblique at $A_{\rm rms}=1$. Now, there are no switchbacks at any point in the evolution. This is simply because of the relationship derived in Sec.~\ref{sec:b0xsc}: according to Eq.~(\ref{eq:b0par}), the normalized amplitude of the parallel (radial in this case) fluctuations is smaller than $A$ by a factor $\sin\theta$. Also visible at late times is the large $b_{0y}$ component: again, this is because $\vp$ rotates towards the $x$-direction, and $\vp\cdot\vb_0=0$ (Eq.~\ref{eq:pb}). Figure~\ref{fig:bxrms} shows the RMS parallel (radial, since $\avg{\vB}$ is in the $x$-direction) component of the fluctuation, $b_{0x,\mathrm{rms}}$ as a function of $A_{\rm rms}$. At small amplitudes, $b_{0x,\mathrm{rms}}\sim A_{\rm rms}^2$, while once the amplitude becomes large and $\theta$ decreases, $b_{0x,\mathrm{rms}}\sim A_{\rm rms}\sin\theta$. This agrees with Eq.~(\ref{eq:b0par}) in both limits, providing some validation for the scaling analysis in Sec.~\ref{sec:b0xsc}.

\begin{figure*}
\includegraphics[width=\linewidth]{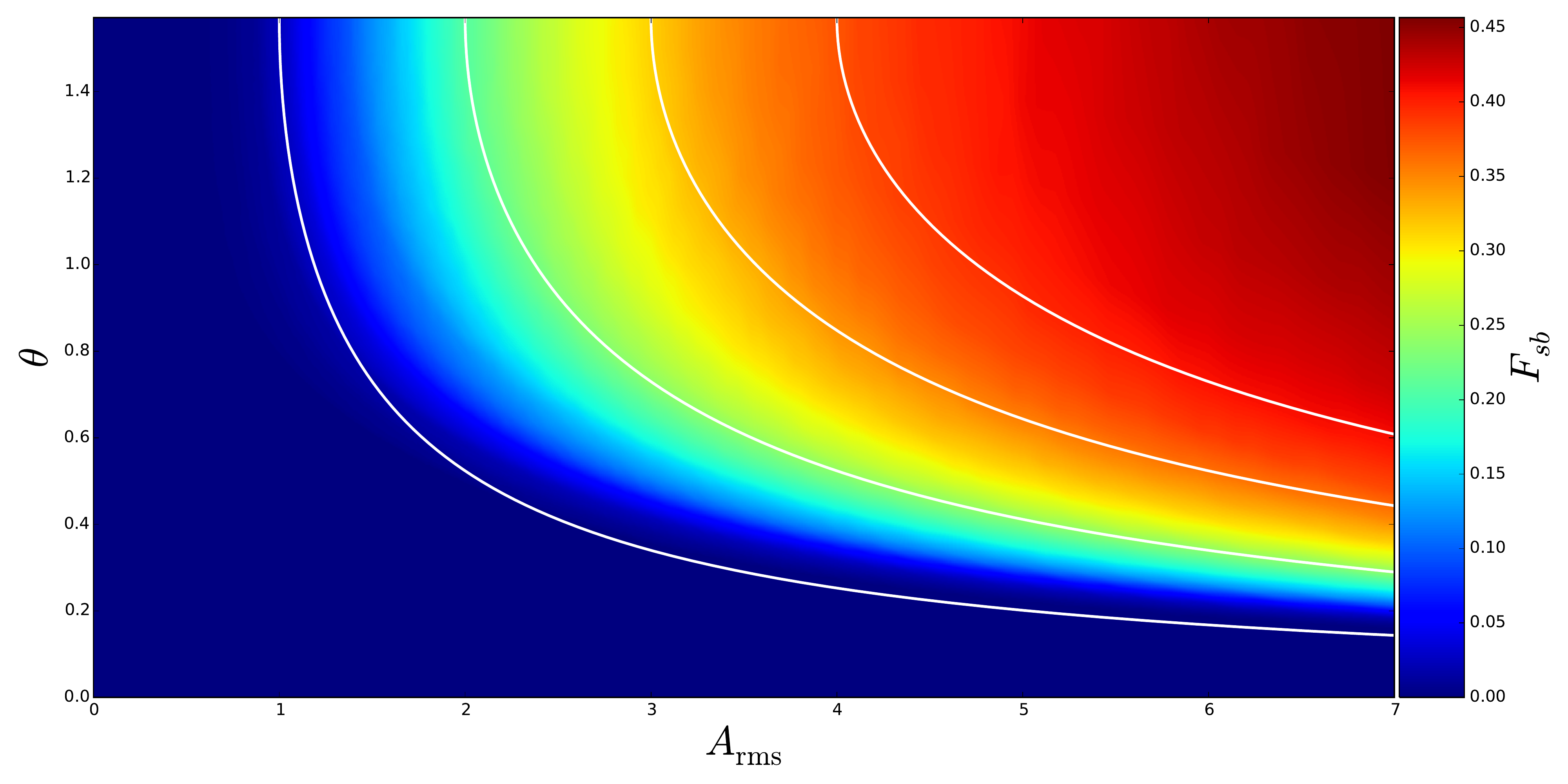}
\caption{The switchback fraction $F_{sb}$ attained by white-noise solutions as a function of $A_{\rm rms}$ and $\theta$. White lines show, from left to right, $A_{\rm rms}\sin\theta=1,2,3,4$ (see Eq.~\ref{eq:b0par}).\label{fig:Athfsb}}
\end{figure*}

To test this scaling analysis more thoroughly, we perform a suite of simulations, with $A_{\rm target}=1$, and vary $\theta_{\rm target}$ from $\pi/8$ to $\pi/2$ in steps of $\pi/400$. We then calculate $F_{\rm sb}(a)$, $A_{\rm rms}(a)$, and $\theta(a)$ for each of these solutions, and interpolate between them in the $A_{\rm rms}$--$\theta$ plane to estimate $F_{\rm sb}(A_{\rm rms},\theta)$. This is shown in Fig.~\ref{fig:Athfsb}, and shows that switchbacks do form more readily for more perpendicular fluctuations: moreover, the contours of $F_{\rm sb}$ roughly agree with the prediction of Eq.~(\ref{eq:b0par}) that the propensity to form radial magnetic-field reversals depends on $A\sin\theta$.\footnote{A detailed analysis shows that the slight difference is caused by the distortion of the waveform; $\mathrm{max}(A)$ scales slightly differently from $A_{\rm rms}$.}
\subsection{Steepening}
\begin{figure}
\begin{center}
\includegraphics[width=\columnwidth]{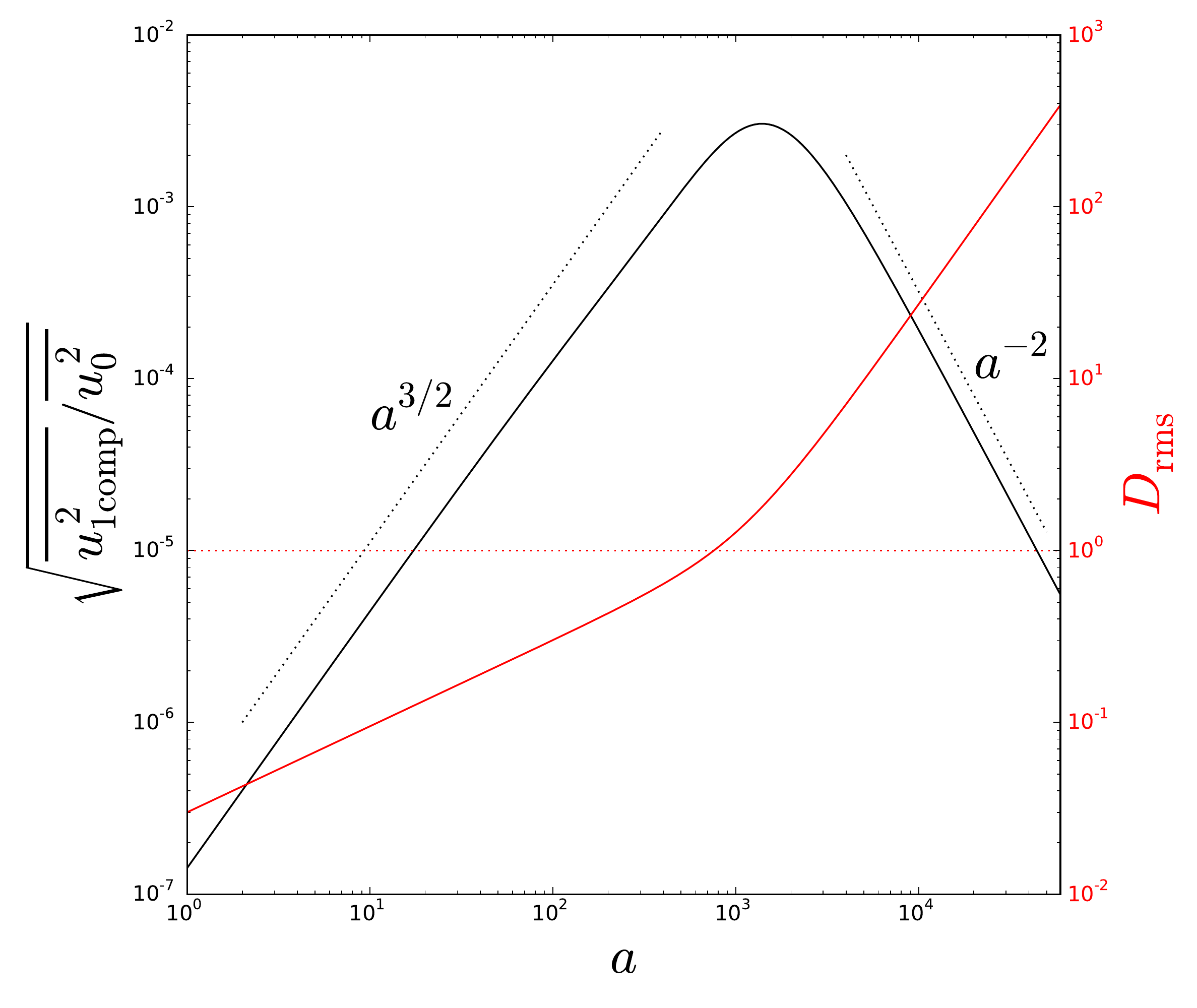}
\end{center}
\caption{Solid black line: RMS $\ucomp/u_0$ as a function of the expansion factor $a$, from the solution whose zeroth-order waveform is shown in Fig.~\ref{fig:lessperp}. Dotted lines show the dependencies in Eq.~(\ref{eq:ucompa}). Solid red line: $D_{\rm rms}$ as a function of $a$ for the same solution. $D_{\rm rms}=1$ is marked as a horizontal red dotted line. \label{fig:ucomp}}
\end{figure}
In Sec.~\ref{sec:vpusc}, we estimated the relative amplitude of the small compressive velocity fluctuations $\ucomp$ and the primary Alfv\'enic velocity fluctuation $u_0$ (Eqs.~\ref{eq:uclowamp} and \ref{eq:uchighamp}). This estimate predicts opposite scalings in the small- and high-amplitude limits, and a maximum where $D=A/\sin\theta\sim 1$ (see Eq.~\ref{eq:Ddef}). We can test this with our simulations. Specifically, we use the simulation with the less perpendicular initialization (\ref{eq:targetlessperp}), whose zeroth-order waveform is shown in Fig.~\ref{fig:lessperp}. The normalized compressive velocity is plotted in Fig.~\ref{fig:ucomp}, and it agrees well with our scaling in both limits. Moreover, the maximum is attained, as expected, when $D_{\rm rms}\sim 1$. This shows that the majority of the steepening occurs when $A_{\rm rms}\sim 1/\sin\theta$, i.e. at moderate normalized amplitude, and at a particular propagation angle. 

\begin{figure*}
\includegraphics[width=\linewidth]{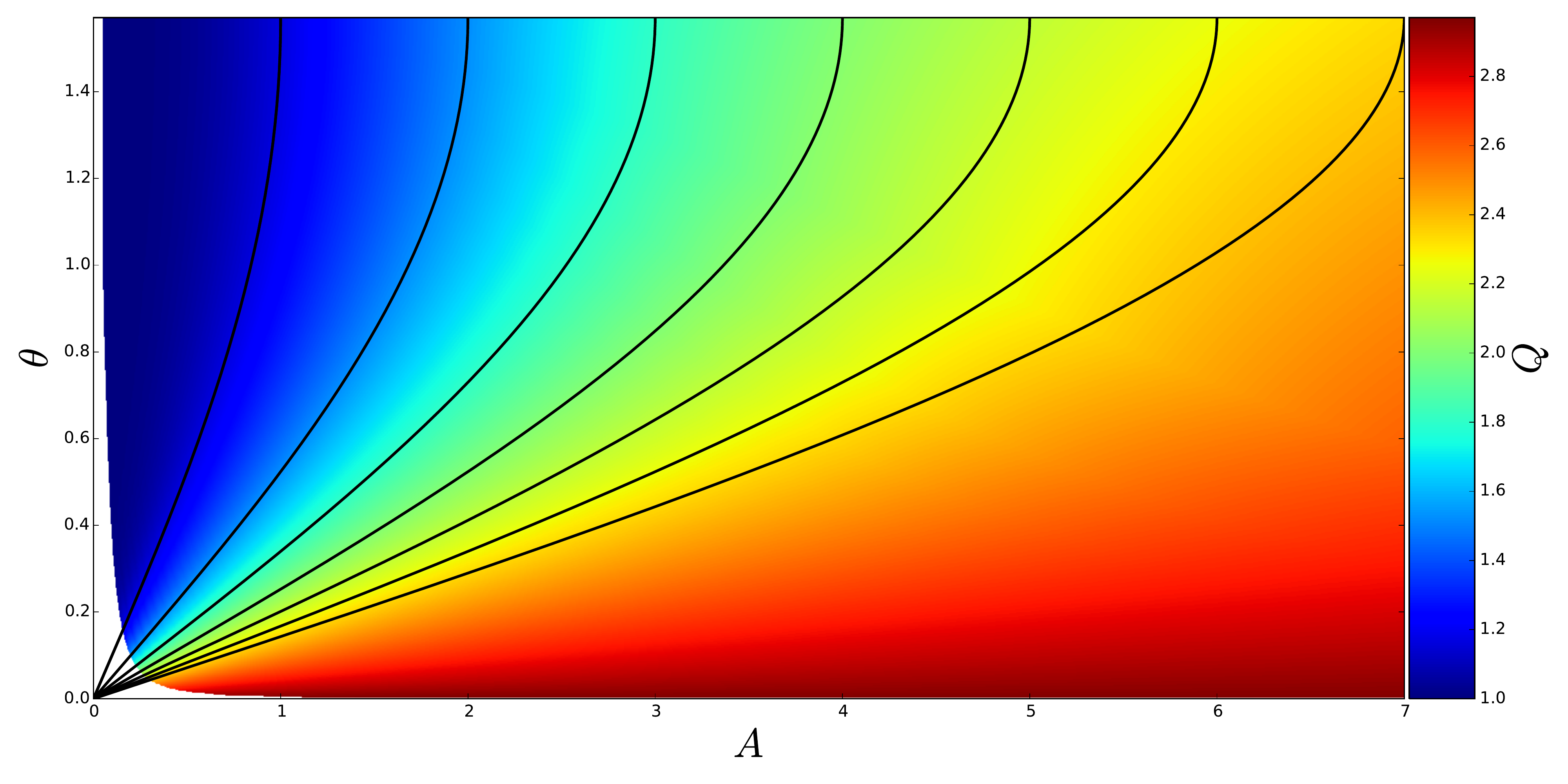}
\caption{The steepness $Q$ (Eq.~\ref{eq:Q}) as a function of $A$ and $\theta$. Black lines show, from left to right, $D=1,2,...,7$. \label{fig:steepness}}
\end{figure*}

A more direct way of looking at the steepening is to examine how the gradients of the primary waveform evolve with time. To do so, we define the ``steepening factor"
\beq
Q= \frac{\avg{|\vb_{0\lambda}|^2}/\avg{|\vb_0|^2}}{\avg{|\vb_{0\lambda}^*|^2}/\avg{|\vb_0^*|^2}},\label{eq:Q}
\eeq
which measures how steep, on average, gradients are compared to the initial condition. We perform a second suite of simulations, now taking
\beq
\theta^* = \arctan\left(\frac{\sin^2\theta_{\rm target}\tan\theta_{\rm target}}{A^{*2}}\right),
\eeq
where $\theta_{\rm target}$ now runs from $\pi/20$ to $\pi/2$ in steps of $\pi/100$. This initialization means that $D_{\rm rms}=1$ (see Eq.~\ref{eq:Ddef}) when $\theta=\theta_{\rm target}$. We calculate $Q(a)$, $A_{\rm rms}(a)$, $\theta(a)$ for each simulation, and interpolate between them in the $A_{\rm rms}$--$\theta$ plane to estimate $Q(A_{\rm rms},\theta)$. This is plotted in Figure~\ref{fig:steepness}. The contours of $Q$, as expected, are lines of constant $D$. This could be tested if a variety of propagation angles could be diagnosed in the data: the large switchbacks appear to be mainly perpendicular, but deflections of the radial field with smaller amplitude could potentially have smaller $\theta$ (and may, according to this analysis, be more steepened). 

However, $Q$ remains quite small for all the simulations. This is in stark contrast to the observed switchbacks (e.g. those in the statistical survey of \citealt{larosa2020}); many of these exhibit extremely steep rotations at their boundaries. Thus, our model cannot explain this property, and additional physics (potentially either due to the effect of three dimensions or kinetics) is required. It should be noted that the switchbacks produced in the MHD turbulence simulations of \citet{squire2020} and \citet{shoda2021} do not appear to be significantly steeper than in our model, which may suggest a kinetic origin to the sharp boundaries observed in the real world; but this could also just be due to the limited resolution available in 3D turbulence simulations.
\subsection{Compressive fluctuations: effect of $\beta$}
\begin{figure*}
\includegraphics[width=\linewidth, trim=0 0.5cm 0 1.2cm, clip]{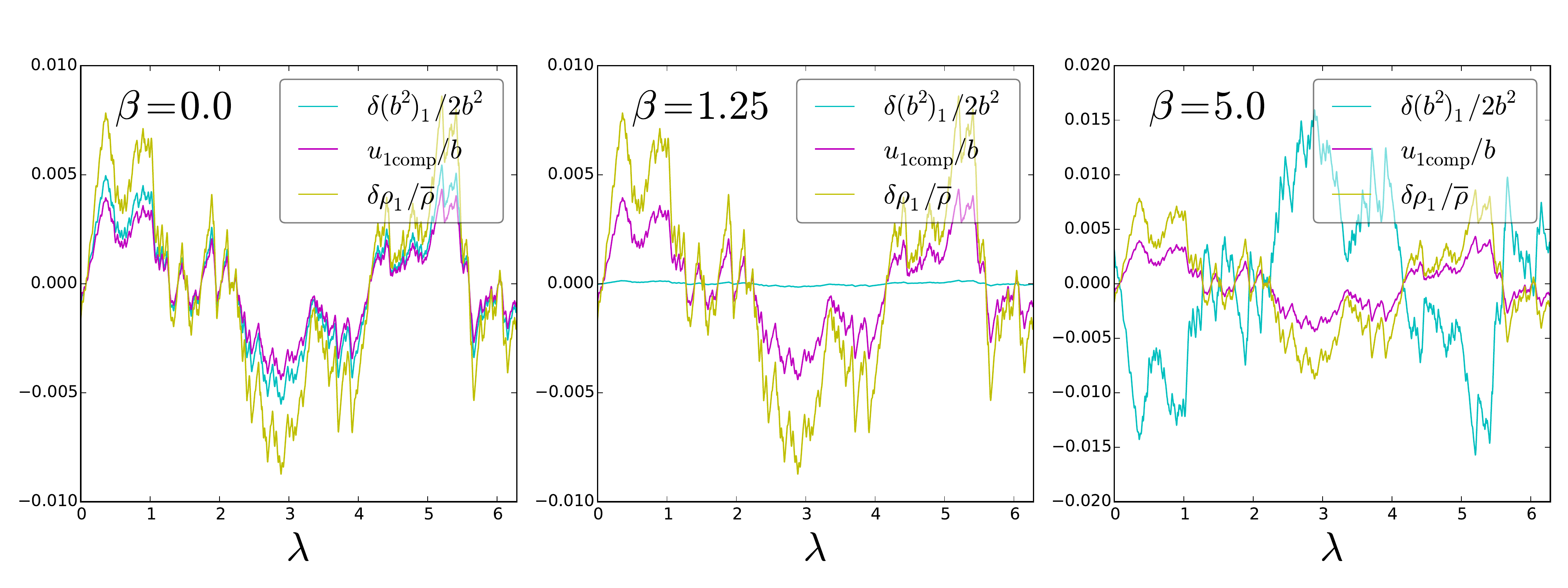}
\caption{The compressive components for the solution shown in the second panel of Fig.~\ref{fig:lessperp}, with $A_{\rm rms} = 1.02$ and $\theta=44.1^\circ$ with $\beta=0,1.25,5$, left to right. Lines show magnetic-field-strength fluctuations $\delta (b^2\!)_1$ (cyan), compressive flow $u_{1\mathrm{comp}}$ (pink), density fluctuations $\delta \rho_1/\rho$ (yellow), with the first two normalized by $b$. \label{fig:beta}}
\end{figure*}

From Eqs.~(\ref{eq:sbeqmain}) and (\ref{eq:vpud}), $\beta$ does not affect the primary Alfv\'en wave at all, only entering through the relationship between $\dbsqm$ and the other variables (see Sec.~\ref{sec:polsc}). For completeness, we plot the compressive fluctuations from the solution in Fig.~\ref{fig:lessperp} when it attains $A=1.02$ and $\theta=44.1^\circ$ at three different $\beta=0,1.25,5$ in Figure~\ref{fig:beta}. Apart from $\dbsq$, the solution does not depend on $\beta$. At low $\beta$, the correlation between $\dbsq$ and $\delta \rho_1$ is positive, but at high $\beta$, the correlation reverses and $\dbsq$ becomes very large relative to the other components, as we deduce in Sec.~\ref{sec:polsc}. Between these two limits, at $\beta=1.25\approx \beta_{\rm crit}$, $\dbsq$ is very small compared to $u_{1comp}$ and $\delta\rho_1$: as expected, for moderate $\theta$, the magnetic-field-strength fluctuations become very small for $\beta\sim1$.

\section{Extensions and further work}\label{sec:extend}
There are several extensions to this work that can and should be performed, relaxing some of the idealized assumptions involved. First, it would be very interesting to go beyond the EBM and investigate the effects of a more realistic expansion model. We plan to investigate the effects of superradial expansion as well as turbulence in a forthcoming work \citep{mallet2021b}.

For the current paper, we have assumed an exactly radial mean magnetic field. Our main dynamical equation (\ref{eq:sbeqmain}) and the expressions for the compressible components (\ref{eq:vpud}--\ref{eq:magpress}) are valid for the mean field in an arbitrary direction, and we plan to investigate the effects of including a Parker spiral (nonzero $\vAy$) in the future. Two interesting effects emerge: first, the degeneracy between the two directions transverse to radial is removed, which could have implications for the observed distribution of switchbacks; second, the transverse components of the mean magnetic field decay more slowly, which will eventually stop the growth in normalized amplitude, as well as rotating the wave back into nearly perpendicular propagation. 

As we mentioned in the introduction, we have not checked the stability of the wave: it is well-known that the Alfv\'en wave is unstable to parametric decay \citep{derby1978,delzanna2001,tenerani2013}. Based on numerical simulations, it appears that this may be somewhat suppressed for localised structures like switchbacks \citep{tenerani2020}, but this process should again be investigated in the future.

Our analysis is formally one-dimensional, and thus it is only truly valid in regions where the gradient direction is not changing too fast, $||\nabla \vp ||/p^2\ll 1$. Perfect Alfv\'enic states of arbitrary amplitude in a homogeneous medium propagate without distortion regardless of their three-dimensional structure, because the group velocity is along $\vA$. However, an attempt to generalize our analysis to three-dimensions causes new terms to appear due to a non-zero $\nabla \vp$: this couples together different directions, and leads to a three-dimensional turbulent cascade. This may lead to more dramatic discontinuities appearing, due to large local fluctuations in $B^2$ (cf. \citealt{cohen1974}, although their analysis only formally applies at small-amplitude and parallel propagation). It is, however, encouraging that some of our scaling results seem to apply to the three-dimensional simulations in \citet{squire2020}. 

We have assumed a locally isothermal equation of state. This does not seem to be sufficient to explain the data, where pressure and density across switchback boundaries have different dependencies \citep{farrell2020}. This limitation may be possible to fix within the framework of MHD, because the compressive components of the wave are small. A more interesting problem would be the case where the pressure can be anisotropic: \citet{squire2016} show that in this case, magnetic-field-strength fluctuations tend to collapse to sharp discontinuities due nullification of the magnetic tension at the boundary. Interestingly, \citet{woodham2020} find that the parallel ion temperature seems to be enhanced in switchback patches, showing that it is probably important to incorporate these effects. Additionally, in the data there is often a proton beam moving at an appreciable fraction of $\vA$; this also alters the local Alfv\'en speed \citep{barnes1971}, and may in fact be driven by phase-steepened Alfv\'en waves themselves \citep{gonzalez2021}: this could also potentially cause additional nonlinear steepening of the wave.

One additional limitation of our approach is that we have assumed without derivation that to leading order the initial condition is spherically polarized to lowest order. However, so long as the initial condition has a small amplitude, this is not too restrictive. For propagation close to parallel ($D\gtrsim 1$ while $A\ll1$), the mechanism of \citet{cohen1974} will apply, and the wave will relax smoothly to an exactly constant-$B^2$ state. For more perpendicular propagation, a small-amplitude fluctuation polarized like a linear Alfv\'en wave may be written as the sum of a small-amplitude constant-$B^2$ Alfv\'en wave and a small-amplitude fast wave \citep{barnes1974}; the latter, due to the significant propagation angle $\theta^*$, will quickly separate from the former (and dissipate due to either steepening into a shock or, in a collisionless plasma, damping): thus, after a short time, only a small-amplitude constant-$B^2$ Alfv\'en wave will remain (this mechanism was first advanced by \citealt{barnes1974} to explain the Alfv\'enicity of the solar wind). Thus, the evolution of small amplitude waves towards a state of constant $B^2$ is reasonably well understood, at least in a homogeneous, non-expanding plasma.  This evolution could be analyzed within a more comprehensive, future version of the model we have developed, but we expect that such a model would lead to conclusions similar to those we have drawn here.

\section{Conclusions}
In this paper, we have developed a basic theory of the evolution of large-amplitude Alfv\'en waves in the expanding solar wind. Our analysis allows us to make several testable predictions, summarized in Table~\ref{tab:pred}. Some of these predictions appear to match hitherto unexplained facets of both spacecraft observations of switchbacks (e.g. \citealt{laker2020,larosa2020,badman2020b}) and numerical simulations of expanding Alfv\'enic turbulence \citep{squire2020}; thus, the physics encoded in our highly idealized setup seems to be relevant in reality.

At a basic level, the physical behaviour encoded in our ``switchback equation" (\ref{eq:sbeqmain}) is as follows: expansion increases the normalized amplitude of the wave (Sec.~\ref{sec:ampsc}), pushing it slightly out of its Alfv\'enic $B^2=\text{constant}$ state. This slight fluctuation in $B^2$ drives a compressive flow, opposing the non-constant-$B^2$, and distorting the waveform to keep the fluctuation in $B^2$ small and under control as the plasma expands. We show that this distortion reaches a maximum when the normalized amplitude $A=\delta B/\avg{B}\approx \sin\theta$, where $\theta$ is the angle between the propagation direction and the mean field $\avg{\vB}$ (Sec.~\ref{sec:vpusc}). However, comparing this distortion with the observed statistical properties of switchbacks (e.g. the relative width of switchback boundaries and cores observed by \citealt{larosa2020}), it does appear that our solutions are not discontinuous enough: thus, there must be additional physics involved; for example, we may need to consider kinetic effects which may enhance steepening due to modifications to the Alfv\'en velocity from pressure anisotropy and/or beams \citep{barnes1971}, or there could be three-dimensional effects that favour the production of more dramatic discontinuities.

We derive analytic expressions for small compressive components of the wave driven by the expansion, in terms of the primary Alfv\'enic component. All these compressive components scale with the expansion rate: this agrees with the turbulence simulations of \citet{squire2020}, who found that higher expansion rates produced larger fluctuations in the magnetic field strength (``magnetic compressions"). Moreover, we show that the normalized density fluctuations (Eq.~\ref{eq:rhoucomp}) are relatively large compared to the normalised compressive velocity when the wave is more perpendicular; this could be tested if the switchbacks can have their direction of propagation diagnosed using a minimum variance analysis. We also show that the small magnetic-field-strength fluctuations vanish at a particular critical $\beta_{\rm crit}$, which generically is of order unity (Sec~\ref{sec:b2dsc}): this hitherto unexplained phenomenon was also observed by \citet{squire2020} in the simulations, and thus it seems that our mechanism for achieving this may be at least somewhat applicable in a more complex and realistic scenario. $\beta_{\rm crit}$ depends on $\theta$, and our prediction for the relationship between $\theta$, $\beta$, and the amplitude of fluctuations in the magnetic field strength could be tested using observational data and/or numerical simulations. Connected with this result, we show that the relative phase between the density and magnetic-field-strength fluctuations reverses at $\beta_{\rm crit}$ (Sec.~\ref{sec:polsc}): this may be connected to the wide range of polarizations observed in real switchbacks by \citet{larosa2020}, and is another testable quantitative prediction. 

We also make predictions regarding the large-amplitude Alfv\'enic (i.e. constant-$B^2$) component of the wave. The overall RMS amplitude scales with the total expansion of the plasma: thus, one might expect patches of switchbacks in regions where the plasma has expanded more \citep{bale2021inprep}, and the switchback fraction should initially increase with radial distance from the sun, as observed \citep{badman2020b,mozer2020}. We show several solutions of Eq.~(\ref{eq:sbeqmain}) in Sec.~\ref{sec:evol}, showing that switchbacks can form from sufficiently large-amplitude Alfv\'en waves, so long as they are sufficiently perpendicular. This preference for perpendicular switchbacks arises simply due to geometry (Sec.~\ref{sec:b0xsc}), and is a consequence of the transverse nature of the fluctuations. This may explain the interesting observation of \citet{laker2020} that switchbacks appear to be highly elongated (by a factor of $10$-$30$) along the magnetic-field direction. Similar physics seems to occur in the numerical simulations: \citet{squire2020} observed that the switchbacks were quite perpendicular structures, and also that initial conditions with more modes close to perpendicular were more prone to generating switchbacks. From our analysis, it appears that the initial condition at the coronal base needs to be spectacularly perpendicular to generate substantial switchbacks. The available data \citep{depontieu2007} does suggest that the initial condition is very perpendicular, but, to match the observed statistics of switchbacks, the evolution of perpendicular and parallel length-scales of the (3D) switchbacks must be significantly different than in our (1D) model. This could potentially arise as a natural consequence of turbulence \citep{gs95}.

In summary, it appears that certain features of the switchbacks can be explained in terms of the nonlinear evolution of Alfv\'en waves as they travel outwards in the expanding solar wind. However, more work is needed to explain the remarkably sharp switchback boundaries in the data, as well as to explain how the switchbacks remain so perpendicular to the mean magnetic field. The main result of our analysis is that although expansion introduces fluctuations in $B^2$ that might naively be expected to destroy the Alfv\'enic state, nonlinearly, these are kept under control by distortion of the primary Alfv\'enic waveform. This provides substantial support for the \citet{squire2020} model of switchback generation, that they are simply the result of the expansion-driven evolution of initially small-amplitude Alfv\'enic fluctuations.

\begin{acknowledgements}
AM would like to thank J. Bonnell, S. Badman, M. McManus, and R. Meyrand for helpful comments and discussions. AM was supported by NASA grant 80NSSC21K0462. Support for JS and RM was provided by Rutherford Discovery Fellowship RDF-U001804 and Marsden Fund grant UOO1727, which are managed through the Royal Society Te Ap\=arangi. BDGC was supported in part by NASA grants NNX17AI18G and 80NSSC19K0829. We acknowledge support from NASA contract NNN06AA01C.
\end{acknowledgements}

\bibliographystyle{aasjournal}
\bibliography{mainbib2}{}

\begin{thebibliography}{}
\expandafter\ifx\csname natexlab\endcsname\relax\def\natexlab#1{#1}\fi
\providecommand{\url}[1]{\href{#1}{#1}}
\providecommand{\dodoi}[1]{doi:~\href{http://doi.org/#1}{\nolinkurl{#1}}}
\providecommand{\doeprint}[1]{\href{http://ascl.net/#1}{\nolinkurl{http://ascl.net/#1}}}
\providecommand{\doarXiv}[1]{\href{https://arxiv.org/abs/#1}{\nolinkurl{https://arxiv.org/abs/#1}}}

\bibitem[{Badman {et~al.}(2020)Badman, Bale, Rouillard, Bowen, Bonnell, Goetz,
  Harvey, MacDowall, Malaspina, \& Pulupa}]{badman2020b}
Badman, S.~T., Bale, S.~D., Rouillard, A.~P., {et~al.} 2020, arXiv preprint
  arXiv:2009.06844

\bibitem[{Bale(2021)}]{bale2021inprep}
Bale, S.D., e.~a. 2021, in preparation

\bibitem[{Bale {et~al.}(2019)Bale, Badman, Bonnell, Bowen, Burgess, Case,
  Cattell, Chandran, Chaston, Chen, {et~al.}}]{bale2019}
Bale, S., Badman, S., Bonnell, J., {et~al.} 2019, Nature, 576, 237

\bibitem[{Barnes \& Hollweg(1974)}]{barnes1974}
Barnes, A., \& Hollweg, J.~V. 1974, J. Geophys. Res., 79, 2302

\bibitem[{Barnes \& Suffolk(1971)}]{barnes1971}
Barnes, A., \& Suffolk, G.~C. 1971, J. Plasma Phys., 5, 315

\bibitem[{{Belcher} \& {Davis}(1971)}]{belcher1971}
{Belcher}, J.~W., \& {Davis}, Jr., L. 1971, J. Geophys. Res., 76, 3534,
  \dodoi{10.1029/JA076i016p03534}

\bibitem[{{Bruno} \& {Carbone}(2013)}]{bruno2013}
{Bruno}, R., \& {Carbone}, V. 2013, Living Rev. Solar Phys., 10, 2,
  \dodoi{10.12942/lrsp-2013-2}

\bibitem[{Chandran(2021)}]{chandran2021}
Chandran, B.~D. 2021, arXiv preprint arXiv:2101.04156

\bibitem[{Chandran \& Perez(2019)}]{chandran2019}
Chandran, B.~D., \& Perez, J.~C. 2019, J. Plasma Phys., 85

\bibitem[{Cohen \& Kulsrud(1974)}]{cohen1974}
Cohen, R.~H., \& Kulsrud, R.~M. 1974, Phys. Fluids, 17, 2215

\bibitem[{Cranmer \& Van~Ballegooijen(2005)}]{cranmer2005}
Cranmer, S., \& Van~Ballegooijen, A. 2005, Astrophys. J. Suppl. Ser., 156, 265

\bibitem[{De~Pontieu {et~al.}(2007)De~Pontieu, McIntosh, Carlsson, Hansteen,
  Tarbell, Schrijver, Shine, Tsuneta, Katsukawa, Ichimoto,
  {et~al.}}]{depontieu2007}
De~Pontieu, B., McIntosh, S., Carlsson, M., {et~al.} 2007, Science, 318, 1574

\bibitem[{de~Wit {et~al.}(2020)de~Wit, Krasnoselskikh, Bale, Bonnell, Bowen,
  Chen, Froment, Goetz, Harvey, Jagarlamudi, {et~al.}}]{dudokdewit2020}
de~Wit, T.~D., Krasnoselskikh, V.~V., Bale, S.~D., {et~al.} 2020, Astrophys. J.
  Suppl. Ser., 246, 39

\bibitem[{Del~Zanna(2001)}]{delzanna2001}
Del~Zanna, L. 2001, Geophys. Res. Lett., 28, 2585

\bibitem[{Derby(1978)}]{derby1978}
Derby, N. 1978, Astrophys. J., 224, 1013

\bibitem[{{Drake} {et~al.}(2020){Drake}, {Agapitov}, {Swisdak}, {Badman},
  {Bale}, {Horbury}, {Kasper}, {MacDowall}, {Mozer}, {Phan}, {Pulupa}, {Szabo},
  \& {Velli}}]{drake2020}
{Drake}, J.~F., {Agapitov}, O., {Swisdak}, M., {et~al.} 2020, arXiv e-prints,
  arXiv:2009.05645.
\newblock \doarXiv{2009.05645}

\bibitem[{Farrell {et~al.}(2020)Farrell, MacDowall, Gruesbeck, Bale, \&
  Kasper}]{farrell2020}
Farrell, W.~M., MacDowall, R.~J., Gruesbeck, J., Bale, S., \& Kasper, J.~C.
  2020, Astrophys. J. Suppl. Ser., 249, 28

\bibitem[{{Goldreich} \& {Sridhar}(1995)}]{gs95}
{Goldreich}, P., \& {Sridhar}, S. 1995, Astrophys. J., 438, 763,
  \dodoi{10.1086/175121}

\bibitem[{Goldstein {et~al.}(1974)Goldstein, Klimas, \& Barish}]{goldstein1974}
Goldstein, M.~L., Klimas, A., \& Barish, F. 1974, Solar Wind III, 385

\bibitem[{Gonz{\'a}lez {et~al.}(2021)Gonz{\'a}lez, Tenerani, Matteini,
  Hellinger, \& Velli}]{gonzalez2021}
Gonz{\'a}lez, C., Tenerani, A., Matteini, L., Hellinger, P., \& Velli, M. 2021,
  arXiv preprint arXiv:2104.02540

\bibitem[{Grappin \& Velli(1996)}]{grappin1996}
Grappin, R., \& Velli, M. 1996, J. Geophys. Res. Space Phys., 101, 425

\bibitem[{Grappin {et~al.}(1993)Grappin, Velli, \& Mangeney}]{grappin1993}
Grappin, R., Velli, M., \& Mangeney, A. 1993, Phys. Rev. Lett., 70, 2190

\bibitem[{Hollweg(1974)}]{hollweg1974}
Hollweg, J.~V. 1974, J. Geophys. Res., 79, 1539

\bibitem[{Horbury {et~al.}(2020)Horbury, Woolley, Laker, Matteini, Eastwood,
  Bale, Velli, Chandran, Phan, Raouafi, {et~al.}}]{horbury2020}
Horbury, T.~S., Woolley, T., Laker, R., {et~al.} 2020, Astrophys. J. Suppl.
  Ser., 246, 45

\bibitem[{Kasper {et~al.}(2019)Kasper, Bale, Belcher, Berthomier, Case,
  Chandran, Curtis, Gallagher, Gary, Golub, {et~al.}}]{kasper2019}
Kasper, J., Bale, S., Belcher, J.~W., {et~al.} 2019, Nature, 576, 228

\bibitem[{Krasnoselskikh {et~al.}(2020)Krasnoselskikh, Larosa, Agapitov,
  de~Wit, Moncuquet, Mozer, Stevens, Bale, Bonnell, Froment,
  {et~al.}}]{krasnoselskikh2020}
Krasnoselskikh, V., Larosa, A., Agapitov, O., {et~al.} 2020, Astrophys. J.,
  893, 93

\bibitem[{Laker {et~al.}(2020)Laker, Horbury, Bale, Matteini, Woolley, Woodham,
  Badman, Pulupa, Kasper, Stevens, {et~al.}}]{laker2020}
Laker, R., Horbury, T.~S., Bale, S.~D., {et~al.} 2020, arXiv preprint
  arXiv:2010.10211

\bibitem[{Larosa {et~al.}(2020)Larosa, Krasnoselskikh, de~Wit, Agapitov,
  Froment, Jagarlamudi, Velli, Bale, Case, Goetz, Harvey, Kasper, Korreck,
  Larson, MacDowall, Malaspina, Pulupa, Revillet, \& Stevens}]{larosa2020}
Larosa, A., Krasnoselskikh, V., de~Wit, T.~D., {et~al.} 2020, Switchbacks:
  statistical properties and deviations from alfv\'enicity.
\newblock \doarXiv{2012.10420}

\bibitem[{Mallet {et~al.}(2021)Mallet, Squire, Chandran, Meyrand, Bowen, \&
  Bale}]{mallet2021b}
Mallet, A., Squire, J., Chandran, B. D.~G., {et~al.} 2021, in preparation

\bibitem[{Matteini {et~al.}(2018)Matteini, Stansby, Horbury, \&
  Chen}]{matteini2018}
Matteini, L., Stansby, D., Horbury, T., \& Chen, C.~H. 2018, Astrophys. J.
  Lett., 869, L32

\bibitem[{McManus {et~al.}(2020)McManus, Bowen, Mallet, Chen, Chandran, Bale,
  Larson, de~Wit, Kasper, Stevens, {et~al.}}]{mcmanus2020}
McManus, M.~D., Bowen, T.~A., Mallet, A., {et~al.} 2020, Astrophys. J. Suppl.
  Ser., 246, 67

\bibitem[{Mozer {et~al.}(2020)Mozer, Agapitov, Bale, Bonnell, Case, Chaston,
  Curtis, de~Wit, Goetz, Goodrich, {et~al.}}]{mozer2020}
Mozer, F., Agapitov, O., Bale, S., {et~al.} 2020, Astrophys. J. Suppl. Ser.,
  246, 68

\bibitem[{Parker(1965)}]{parker1965}
Parker, E. 1965, Space Sci. Rev., 4, 666

\bibitem[{{Perez} \& {Chandran}(2013)}]{perezchandran2013}
{Perez}, J.~C., \& {Chandran}, B.~D.~G. 2013, Astrophys. J., 776, 124,
  \dodoi{10.1088/0004-637X/776/2/124}

\bibitem[{{Ruffolo} {et~al.}(2020){Ruffolo}, {Matthaeus}, {Chhiber}, {Usmanov},
  {Yang}, {Bandyopadhyay}, {Parashar}, {Goldstein}, {DeForest}, {Wan},
  {Chasapis}, {Maruca}, {Velli}, \& {Kasper}}]{ruffolo2020}
{Ruffolo}, D., {Matthaeus}, W.~H., {Chhiber}, R., {et~al.} 2020, arXiv
  e-prints, arXiv:2009.06537.
\newblock \doarXiv{2009.06537}

\bibitem[{Schwadron \& McComas(2021)}]{schwadron2021}
Schwadron, N., \& McComas, D. 2021, arXiv preprint arXiv:2102.03696

\bibitem[{Shoda {et~al.}(2021)Shoda, Chandran, \& Cranmer}]{shoda2021}
Shoda, M., Chandran, B.~D., \& Cranmer, S.~R. 2021, arXiv preprint
  arXiv:2101.09529

\bibitem[{Squire {et~al.}(2020)Squire, Chandran, \& Meyrand}]{squire2020}
Squire, J., Chandran, B.~D., \& Meyrand, R. 2020, Astrophys. J. Lett., 891, L2

\bibitem[{Squire {et~al.}(2016)Squire, Quataert, \& Schekochihin}]{squire2016}
Squire, J., Quataert, E., \& Schekochihin, A. 2016, Astrophys. J. Lett., 830,
  L25

\bibitem[{Tenerani \& Velli(2013)}]{tenerani2013}
Tenerani, A., \& Velli, M. 2013, J. Geophys. Res. Space Phys., 118, 7507

\bibitem[{Tenerani {et~al.}(2020)Tenerani, Velli, Matteini, R{\'e}ville, Shi,
  Bale, Kasper, Bonnell, Case, de~Wit, {et~al.}}]{tenerani2020}
Tenerani, A., Velli, M., Matteini, L., {et~al.} 2020, Astrophys. J. Suppl.
  Ser., 246, 32

\bibitem[{Van~Ballegooijen \& Asgari-Targhi(2016)}]{ballegooijen2016}
Van~Ballegooijen, A., \& Asgari-Targhi, M. 2016, Astrophys. J., 821, 106

\bibitem[{van Ballegooijen \& Asgari-Targhi(2017)}]{ballegooijen2017}
van Ballegooijen, A.~A., \& Asgari-Targhi, M. 2017, Astrophys. J., 835, 10

\bibitem[{Vasquez \& Hollweg(1998)}]{vasquez1998b}
Vasquez, B.~J., \& Hollweg, J.~V. 1998, J. Geophys. Res. Space Phys., 103, 335

\bibitem[{{Velli} {et~al.}(1992){Velli}, {Grappin}, \& {Mangeney}}]{velli1992}
{Velli}, M., {Grappin}, R., \& {Mangeney}, A. 1992, in American Institute of
  Physics Conference Series, Vol. 267, Electromechanical Coupling of the Solar
  Atmosphere, ed. D.~S. {Spicer} \& P.~{MacNeice}, 154--159,
  \dodoi{10.1063/1.42861}

\bibitem[{Verdini \& Velli(2007)}]{verdini2007}
Verdini, A., \& Velli, M. 2007, Astrophys. J., 662, 669

\bibitem[{Wang \& Sheeley(1990)}]{wang1990}
Wang, Y.-M., \& Sheeley, N. 1990, Astrophys. J., 355, 726

\bibitem[{Woodham {et~al.}(2020)Woodham, Horbury, Matteini, Woolley, Laker,
  Bale, Nicolaou, Stawarz, Stansby, Hietala, {et~al.}}]{woodham2020}
Woodham, L., Horbury, T., Matteini, L., {et~al.} 2020, arXiv preprint
  arXiv:2010.10379

\bibitem[{Zank {et~al.}(2020)Zank, Nakanotani, Zhao, Adhikari, \&
  Kasper}]{zank2020}
Zank, G., Nakanotani, M., Zhao, L.-L., Adhikari, L., \& Kasper, J. 2020,
  Astrophys. J., 903, 1

\end{thebibliography}

\end{document}